\documentclass[twocolumn]{aastex631}

\usepackage{subfigure}
\usepackage{url}
\usepackage{hyperref}
\usepackage{epsfig}
\usepackage{graphicx}
\usepackage{sidecap}
\usepackage{hanging}
\usepackage{color}
\usepackage{multirow}
\usepackage{enumerate}
\usepackage{natbib}
\usepackage{amssymb}
\usepackage{amsmath}
\usepackage[bottom]{footmisc}

\newcommand{\sci}{Science}
\newcommand{\jatis}{JATIS}

\definecolor{steelblue}{rgb}{0.0, 0.0, 0.5}
\definecolor{brightmaroon}{rgb}{0.76, 0.13, 0.28}

\newcommand{\kepler}{{Kepler}}

\providecommand{\e}[1]{\ensuremath{\, \times \, 10^{#1}}}

\providecommand{\bjdtdb}{\ensuremath{\rm {BJD_{TDB}}}}

\providecommand{\msun}{\ensuremath{\,M_\Sun}}
\providecommand{\rsun}{\ensuremath{\,R_\Sun}}
\providecommand{\lsun}{\ensuremath{\,L_\Sun}}
\providecommand{\mj}{\ensuremath{\,M_{\rm J}}}
\providecommand{\rj}{\ensuremath{\,R_{\rm J}}}

\providecommand{\fave}{\langle F \rangle}
\providecommand{\fluxcgs}{10$^9$ erg s$^{-1}$ cm$^{-2}$}

\accepted{December 20, 2023}

\submitjournal{ApJS}

\shorttitle{GOT `EM IV. Masses \& Radii of Long-period Giant Exoplanets}
\shortauthors{P. Dalba et al.}

\begin{document}

\title{Giant Outer Transiting Exoplanet Mass (GOT `EM) Survey. IV.\\Long-term Doppler Spectroscopy for 11 Stars Thought to Host Cool Giant Exoplanets}

\correspondingauthor{Paul A. Dalba}
\email{pauldalba.astro@gmail.com}


\author[0000-0002-4297-5506]{Paul A.\ Dalba} 
\altaffiliation{Heising-Simons 51 Pegasi b Postdoctoral Fellow}
\affiliation{Department of Astronomy and Astrophysics, University of California, Santa Cruz, CA 95064, USA}
\affiliation{SETI Institute, Carl Sagan Center, 339 Bernardo Ave, Suite 200, Mountain View, CA 94043, USA}

\author[0000-0002-7084-0529]{Stephen R.\ Kane} 
\affiliation{Department of Earth and Planetary Sciences, University of California, Riverside, CA 92521, USA}

\author[0000-0002-0531-1073]{Howard Isaacson} 
\affiliation{{Department of Astronomy,  University of California Berkeley, Berkeley CA 94720, USA}}
\affiliation{Centre for Astrophysics, University of Southern Queensland, Toowoomba, QLD, Australia}

\author[0000-0003-3504-5316]{Benjamin Fulton} 
\affiliation{NASA Exoplanet Science Institute/Caltech-IPAC, MC 314-6, 1200 E. California Blvd., Pasadena, CA 91125, USA}

\author[0000-0001-8638-0320]{Andrew W.\ Howard} 
\affiliation{Department of Astronomy, California Institute of Technology, Pasadena, CA 91125, USA}

\author[0000-0002-2949-2163]{Edward W.\ Schwieterman} 
\affiliation{Department of Earth and Planetary Sciences, University of California, Riverside, CA 92521, USA}

\author[0000-0002-5113-8558]{Daniel P.\ Thorngren} 
\affiliation{Department of Physics \& Astronomy, Johns Hopkins University, Baltimore, MD, USA}

\author[0000-0002-9843-4354]{Jonathan Fortney} 
\affiliation{Department of Astronomy and Astrophysics, University of California, Santa Cruz, CA 95064, USA}

\author[0000-0002-0701-4005]{Noah Vowell} 
\affiliation{Center for Data Intensive and Time Domain Astronomy, Department of Physics and Astronomy, Michigan State University, East Lansing, MI 48824, USA}

\author[0000-0001-7708-2364]{Corey Beard} 
\altaffiliation{NASA FINESST Fellow}
\affiliation{Department of Physics \& Astronomy, University of California Irvine, Irvine, CA 92697, USA}

\author[0000-0002-3199-2888]{Sarah Blunt} 
\affiliation{Department of Astronomy, California Institute of Technology, Pasadena, CA 91125, USA}

\author[0000-0002-4480-310X]{Casey L. Brinkman} 
\affiliation{Institute for Astronomy, University of Hawai'i, 2680 Woodlawn Drive, Honolulu, HI 96822 USA}

\author[0000-0003-1125-2564]{Ashley Chontos} 
\altaffiliation{Henry Norris Russell Fellow}
\affiliation{Department of Astrophysical Sciences, Princeton University, 4 Ivy Lane, Princeton, NJ 08540, USA}
\affiliation{Institute for Astronomy, University of Hawai`i, 2680 Woodlawn Drive, Honolulu, HI 96822, USA} 

\author[0000-0002-8958-0683]{Fei Dai} 
\altaffiliation{NASA Sagan Fellow}
\affiliation{Division of Geological and Planetary Sciences,
1200 E California Blvd, Pasadena, CA, 91125, USA}
\affiliation{Department of Astronomy, California Institute of Technology, Pasadena, CA 91125, USA}

\author[0000-0002-8965-3969]{Steven Giacalone} 
\affil{Department of Astronomy, University of California Berkeley, Berkeley, CA 94720, USA} 

\author[0000-0002-0139-4756]{Michelle L. Hill} 
\affiliation{Department of Earth and Planetary Sciences, University of California, Riverside, CA 92521, USA}

\author[0000-0002-6115-4359]{Molly Kosiarek} 
\affiliation{Department of Astronomy and Astrophysics, University of California, Santa Cruz, CA 95064, USA}

\author[0000-0001-8342-7736]{Jack Lubin} 
\affiliation{Department of Physics \& Astronomy, University of California Irvine, Irvine, CA 92697, USA}

\author[0000-0002-7216-2135]{Andrew W. Mayo} 
\affiliation{Department of Astronomy, University of California Berkeley, Berkeley CA 94720, USA} 
\affiliation{Centre for Star and Planet Formation, Natural History Museum of Denmark \& Niels Bohr Institute, University of Copenhagen, \O ster Voldgade 5-7, DK-1350 Copenhagen K., Denmark}

\author[0000-0003-4603-556X]{Teo Mo\v{c}nik} 
\affiliation{Gemini Observatory/NSF's NOIRLab, 670 N. A'ohoku Place, Hilo, HI 96720, USA}

\author[0000-0001-8898-8284]{Joseph M. Akana Murphy} 
\altaffiliation{NSF Graduate Research Fellow}
\affiliation{Department of Astronomy and Astrophysics, University of California, Santa Cruz, CA 95064, USA}

\author[0000-0003-0967-2893]{Erik A. Petigura} 
\affiliation{Department of Physics \& Astronomy, University of California Los Angeles, Los Angeles, CA 90095, USA} 

\author[0000-0002-7670-670X]{Malena Rice} 
\affiliation{Department of Astronomy, Yale University, New Haven, CT 06511, USA}

\author[0000-0003-3856-3143]{Ryan A. Rubenzahl} 
\altaffiliation{NSF Graduate Research Fellow}
\affiliation{Department of Astronomy, California Institute of Technology, Pasadena, CA 91125, USA}
 
\author[0000-0002-4290-6826]{Judah Van Zandt} 
\affiliation{Department of Physics \& Astronomy, University of California Los Angeles, Los Angeles, CA 90095, USA}

\author[0000-0002-3725-3058]{Lauren M. Weiss} 
\affiliation{Department of Physics and Astronomy, University of Notre Dame, Notre Dame, IN 46556, USA}

\author[0000-0003-2313-467X]{Diana Dragomir} 
\affiliation{Department of Physics \& Astronomy, University of New Mexico, 1919 Lomas Blvd NE, Albuquerque, NM 87131, USA}

\author[0000-0002-4365-7366]{David Kipping} 
\affiliation{Dept. of Astronomy, Columbia University, 550 W 120th Street, New York NY 10027}

\author[0000-0001-5133-6303]{Matthew J.\ Payne} 
\affiliation{Harvard-Smithsonian Center for Astrophysics, 60 Garden St., MS 51, Cambridge, MA 02138, USA}

\author[0000-0001-8127-5775]{Arpita Roy} 
\affiliation{Space Telescope Science Institute, Baltimore, MD 21218, USA}
\affiliation{Department of Physics and Astronomy, Johns Hopkins University, Baltimore, MD 21218, USA}

\author[0000-0003-2331-5606]{Alex Teachey} 
\affiliation{Academia Sinica Institute of Astronomy \& Astrophysics, Taipei, Taiwan R.O.C.}

\author[0000-0001-6213-8804]{Steven Villanueva Jr.} 
\affiliation{NASA Goddard Space Flight Center, 8800 Greenbelt Road, Greenbelt, MD 20771, USA}


\begin{abstract}
Discovering and characterizing exoplanets at the outer edge of the transit method's sensitivity has proven challenging owing to geometric biases and the practical difficulties associated with acquiring long observational baselines. Nonetheless, a sample of giant exoplanets on orbits longer than 100 days has been identified by transit hunting missions. We present long-term Doppler spectroscopy for 11 such systems with observation baselines spanning a few years to a decade. We model these radial velocity observations jointly with transit photometry to provide initial characterizations of these objects and the systems in which they exist. Specifically, we make new precise mass measurements for four long-period giant exoplanets (Kepler-111 c, Kepler-553 c, Kepler-849 b, and PH-2 b), we place new upper limits on mass for four others (Kepler-421 b, KOI-1431.01, Kepler-1513 b, and Kepler-952 b), and we show that several ``confirmed'' planets are in fact not planetary at all. We present these findings to complement similar efforts focused on closer-in short-period giant planets, and with the hope of inspiring future dedicated studies of cool giant exoplanets.
\end{abstract}


\section{Introduction} \label{sec:intro}

The orbits of extrasolar planets are not preferentially inclined for our vantage point from the Solar System. Whether we are positioned properly to observe an exoplanet transit is essentially up to chance. However, this chance is far from uniform across all exoplanets. To first order, the transit probability is the ratio between the stellar radius and the planet's semi-major axis. Planets with close-in (i.e., short-period) orbits are more likely to transit their hosts from our point of view. With larger semi-major axes, the transit probability falls off accordingly. In a practical sense, the transit detection probability is even more strongly biased against long-period planets \citep[e.g.,][]{Beatty2008}. Given that the timing of conjunction across all exoplanet systems is random, transit surveys must search for long-period planets with similar long baselines of observation.  

An observation of an exoplanet in transit provides a wealth of scientific information. Though indirectly, transits can reveal a planet's size, orbital characteristics, and even its atmospheric properties \citep[e.g.,][]{Seager2000,Seager2003}. When coupled with a dynamical mass measurement, for which a transiting geometry is also favorable, the information stemming from a transit effectively completes the first-order characterization of the transiting object. Overall, transits have played (and continue to play) a fundamental role in the development of exoplanet science. As such, there is strong scientific motivation to overcome the statistical and practical challenges associated with observing the longest-period transiting exoplanets.

Probing the outer edge of the transit method's sensitivity has proven challenging. Ground-based exoplanet detection efforts are generally fruitless for exoplanets with orbits longer than a few weeks \citep[e.g.,][]{Pollacco2006,Pepper2007,Brahm2016}. Space-based facilities generally achieve longer baselines and have more success at finding exoplanets on orbits of dozens to hundreds of days. The Kepler mission \citep{Borucki2010} remains as the gold standard in this regard, and its four-year staring strategy discovered some of the longest-period transiting exoplanets known to exist \citep{Kipping2016b,Dalba2021c}. The limitations of the Kepler mission (e.g., faint host stars, narrow field of view) are being overcome by the ongoing Transiting Exoplanet Survey Satellite (TESS) mission, although at the cost of observational baseline \citep{Ricker2015}. TESS systematically covers the entire sky with $\sim28$-day ``sectors.'' However, as the TESS mission continues, it has slowly but surely built up baselines and discovered long-period planets akin to those spotted by Kepler \citep[e.g.,][]{Kostov2021,Dalba2022a,Heitzmann2023,OrellMiquel2023,Mann2023,Mireles2023} but orbiting far brighter host stars. 

The discovery of transits by a long-period exoplanet candidate, as difficult and unlikely as it may be, is just the beginning of the challenge. Follow-up observations, specifically Doppler spectroscopy to measure the object's orbital elements and mass, typically require several orbits worth of observations. Long-term radial velocity (RV) surveys have proven very successful \citep[e.g.,][]{Mayor2011,Wittenmyer2020,Rosenthal2021}, but most either started before the era of transiting exoplanets or were constructed around other goals. A notable exception is \citet{Santerne2016}, which followed up Kepler giant planets having orbital periods less than 400~days with the SOPHIE spectrograph on a 1.9~m telescope. The combination of long periods and low masses, even only down to a Saturn mass, however, necessitates a larger aperture to achieve better RV precision and more confident mass measurements. Moreover, an RV program for long-period transiting exoplanets requires long-term but low cadence observations, ideally with flexibility to account for orbital eccentricities, undetected companions, etc. Accessing a large aperture facility in this manner can present practical challenges that further hinder long-period exoplanet discovery and characterization.

Here we present the next installment from the Giant Outer Transiting Exoplanet Mass (GOT `EM) survey. The GOT `EM survey aims to provide long-term RV follow up to the longest-period transiting giant planets discovered by Kepler or TESS \citep{Dalba2020b,Dalba2021a,Dalba2021c,Mann2023}. In what follows, we publish new RV data from multiple facilities and either precise masses or upper limits for the following systems thought to contain transiting giant exoplanets with orbital periods greater than 100~days: Kepler-111 c (KOI-139.01), Kepler-553~c (KOI-443.02), Kepler-421~b (KOI-1274.01), Kepler-807~b (KOI-1288.01), KOI-1431.01, Kepler-849~b (KOI-1439.01), Kepler-952~b (KOI-1790.01), PH-2~b (KOI-3663.01), Kepler-1513~b (KOI-3678.01), KIC~5951458~b (Kepler-456~b), and TOI-2180~b. In Section~\ref{sec:obs}, we describe the collection and treatment of the photometric and spectroscopic data. In Section~\ref{sec:methods}, we outline the general analysis procedure applied to each system. In Section~\ref{sec:res}, we present our findings for each system individually and mention any special treatments applied during its analysis. In total, we publish three new mass measurements, four new mass upper limits, and four mass updates for these systems as well as mass and orbital property refinements of any non-transiting outer companions. In Section~\ref{sec:disc}, we discuss the importance of several of the individual systems and the work as a whole. Finally, in the Appendix, we provide extensive tables of stellar and planetary properties for a subset of the planetary systems studied here.


\section{Observations} \label{sec:obs}


\subsection{Kepler Photometry} \label{sec:kepler}

For all systems except TOI-2180, we accessed the Kepler spacecraft photometry from the Mikulski Archive for Space Telescopes (MAST) using the \textsf{lightkurve} package \citep{Lightkurve2018}. We identified quarters containing transits using each system's data validation report. If short and long cadence data were both present for a quarter containing a transit, we used the short cadence data by default. In all cases, we used the Pre-search Data Conditioning Simple Aperture Photometry \citep[PDCSAP;][]{Jenkins2010a,Smith2012,Stumpe2012} and also applied our own spline detrending using \textsf{keplerspline} \citep{Vanderburg2016b}. This procedure produced flattened transit light curves for all systems. Transit timing variations (TTVs) are handled on a case-by-case basis as described for relevant systems in Section~\ref{sec:res}. In the case of Kepler-553, we followed these same steps for transits of the inner (b) planet as well.


\subsection{TESS Photometry} \label{sec:tess}

We made use of photometric observations from TESS for one target: TOI-2180~b. Two transits of this planet had previously been published \citep{Dalba2022a,Dalba2022b}. We adopted the fully processed data from those studies and then access one additional unpublished transit that TESS observed in Sector~57. These observations were conducted in fast (20~second) cadence. We downloaded this transit data from MAST using the \textsf{lightkurve} package \citep{Lightkurve2018}. We accessed the PDCSAP flux, which had been filtered for high frequency noise. Following the same procedure as \citet{Dalba2022b} for the first two transits, we then used \textsf{keplerspline} to further flatten the transit data \citep{Vanderburg2016b}. 


\subsection{Keck-HIRES Spectroscopy} \label{sec:hires}

The High Resolution Echelle Spectrometer\citep[HIRES;][]{Vogt1994} on the Keck~I telescope has a rich history of making stable RV measurements of exoplanet host stars over decades long timescales \citep[e.g.,][]{Rosenthal2021}. We made use of this facility's ability and achieved RV baselines ranging from a few years to a decade for all of our systems. 

RVs were extracted via the iodine technique whereby star light is passed through a heated I$_2$ cell that imprints reference lines on the stellar spectrum \citep[e.g.,][]{Butler1996}. These spectra are then compared with a high signal-to-noise ratio (S/N) template spectrum that was acquired without the I$_2$ cell via a forward modeling procedure that yields the final RV \citep{Howard2010,Howard2016}. Table~\ref{tab:rv_hires} contains all of the Keck-HIRES RVs used in this study except for those of TOI-2180, which are provided in Table~\ref{tab:rv_toi2180}. 

\begin{deluxetable*}{ccccccc}
\tabletypesize{\footnotesize}
\tablecaption{RV Measurements of Kepler Planet Hosting Stars From Keck-HIRES. \label{tab:rv_hires}}
\tablehead{
  \colhead{KIC} &
  \colhead{KOI} & 
  \colhead{Kepler} &
  \colhead{BJD$_{\rm TDB}$} & 
  \colhead{RV (m s$^{-1}$)\tablenotemark{a}} &
  \colhead{$S_{\rm HK}$} &
  \colhead{Template\tablenotemark{b}}
  }
\startdata
8559644 & 139 & 111 & 2455669.098798 & $-17.1\pm5.7$ & $0.140\pm0.001$ & Match \\
8559644 & 139 & 111 & 2455703.948116 & $-31.8\pm6.2$ & $0.110\pm0.001$ & Match \\
8559644 & 139 & 111 & 2455782.996098 & $22.0\pm3.3$ & $0.147\pm0.001$ & Match \\
8559644 & 139 & 111 & 2457207.989656 & $8.6\pm3.1$ & $0.133\pm0.001$ & Match \\
8559644 & 139 & 111 & 2459739.998472 & $-14.1\pm2.9$ & $0.148\pm0.001$ & Match \\
8559644 & 139 & 111 & 2459771.936477 & $-5.9\pm3.4$ & $0.156\pm0.001$ & Match \\
8559644 & 139 & 111 & 2459822.905135 & $22.6\pm3.4$ & $0.157\pm0.001$ & Match \\
8559644 & 139 & 111 & 2460107.980866 & $14.5\pm2.9$ & $0.138\pm0.001$ & Match \\
10937029 & 433 & 553 & 2458387.933620 & $-34.8\pm8.8$ & $0.174\pm0.001$ & Match \\
10937029 & 433 & 553 & 2458392.909029 & $-84.2\pm7.4$ & $0.265\pm0.001$ & Match \\
\enddata
\tablenotetext{}{This is a representative subset of the full data set. The full table will be made available in machine readable format.}
\tablenotetext{a}{These values are relative RVs. The systematic velocity offset has been removed.}
\tablenotetext{b}{``Self'' indicates that a template of the same host star was used to generate the RVs. ``Match'' indicates that a best match template was used and future modeling effort with these RVs should include a jitter term (see Section~\ref{sec:hires}).}
\end{deluxetable*}

There are two important points to make regarding the Keck-HIRES RVs published here. First, in some cases, the template spectrum used in the RV extraction procedure is not that of the host star from which the I$_2$ spectra have been collected. For the faintest host stars, we swap in a ``best match'' template following \citet{Dalba2020b}. This technique provides for precise RV extraction, though typically incurs some jitter, which we allow for in our fits. The style of template used for each observation is specified in Table~\ref{tab:rv_hires}. Second, the RV extraction procedure is applied to all available data simultaneously. Consequently, after each new spectrum is collected, all of the inferred RV values change. As more data are collected, the change in RV typically becomes increasingly negligible. The timestamps of the data do not change. For the two cases in which we make use of previously published Keck-HIRES RVs (KIC~5951458 and TOI-2180), we publish all of the RVs to allow for consistent reproduction of this work. 

\begin{figure*}
    \centering
    \includegraphics[width=\textwidth]{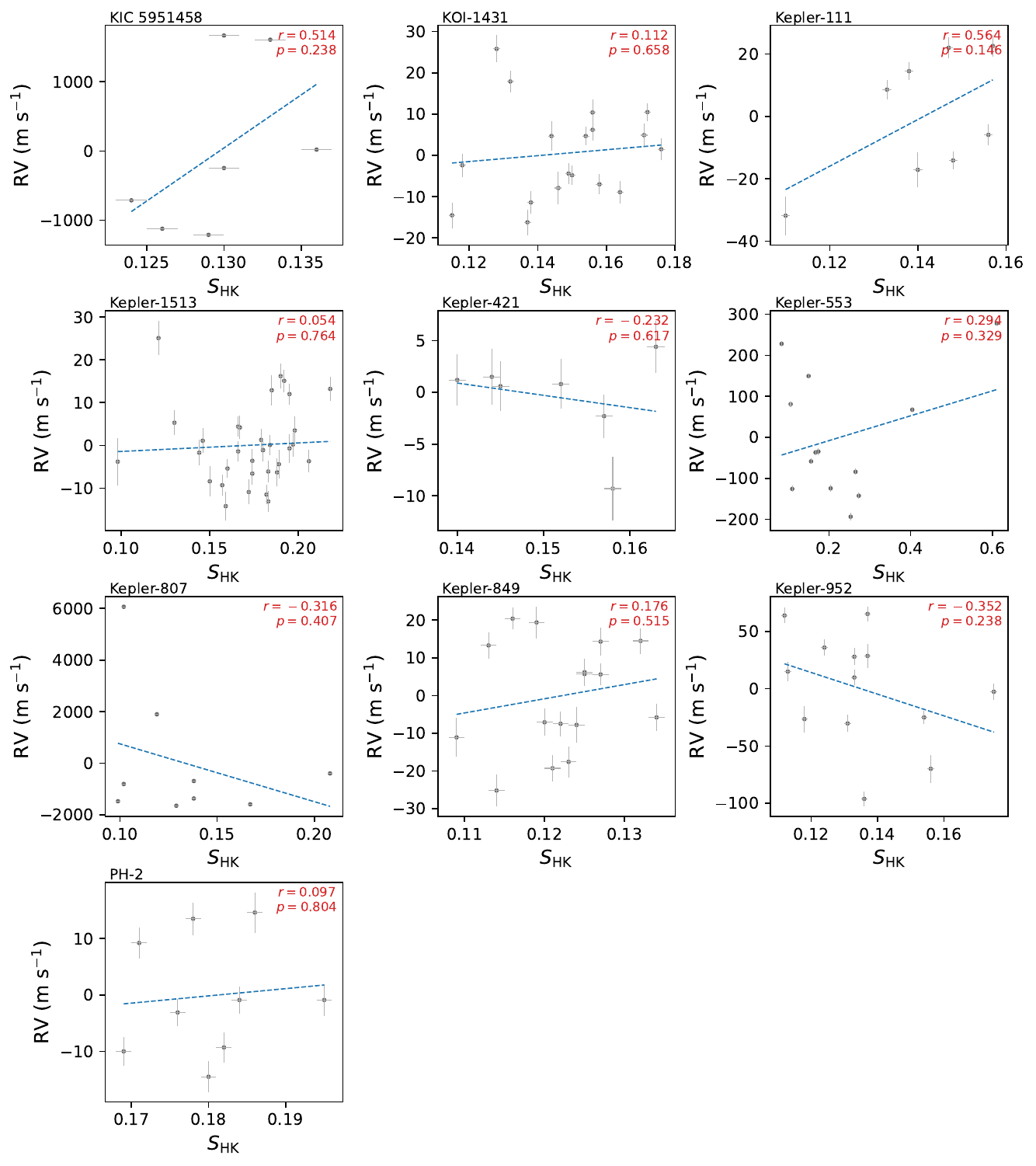}
    \caption{RV measurements versus $S_{\rm HK}$ activity indicators for all of the Kepler targets. The blue dashed line is a linear regression between these parameters that is described by the correlation coefficient ($r$) and the $p$-value displayed in each panel. None of the systems show significant correlation between RV and stellar activity.}
    \label{fig:svalue}
\end{figure*}

Tables~\ref{tab:rv_hires} and \ref{tab:rv_toi2180} also contain the $S_{\rm HK}$ stellar activity indicator derived from the Ca II H and K spectral lines measured in the I$_2$ spectra \citep{Wright2004,Isaacson2010}. Figure~\ref{fig:svalue} displays these activity indicators alongside their corresponding RV values. By basic linear regression, we do not identify statistically significant correlation between the $S_{\rm HK}$ and RV values for any of our systems.

In addition to I$_2$ and template spectra, we also acquired a low S/N spectrum for each system that was used for basic stellar characterization and vetting against false positive scenarios. If a template spectrum of the actual host star was later acquired, it was then used for the stellar characterization. The stellar characterization analysis was conducted with \textsf{SpecMatch} \citep{Petigura2015,Petigura2017b}, which returned spectroscopically derived estimates of stellar effective temperature ($T_{\rm eff}$), iron abundance ([Fe/H]), and surface gravity ($\log g$). 

The false positive analysis was conducted with \textsf{ReaMatch} \citep{Kolbl2015}, which searches for evidence of a second set of spectral lines as would be seen in the case of an SB2. \textsf{ReaMatch} calculates the cross correlation function between spectra of the target star and a similar star drawn from the HIRES library. It also conducts a similar analysis on the residuals of the target star's and best match star's spectra. In all cases, these cross correlation functions only peaked at $\Delta$RV=0~m~s$^{-1}$, indicating that only a single set of spectral lines was present in each spectrum down to the $\sim$1\% relative flux level.


\subsection{NEID Spectroscopy} \label{sec:neid}

We collected five spectra for Kepler-1513 using the NEID spectrograph on the 3.5~m WIYN telescope at Kitt Peak Observatory\footnote{WIYN is a joint facility of the University of Wisconsin Madison, Indiana University, the NSF's NOIRLab, the Pennsylvania State University, Purdue University, the University of California  Irvine, and the University of Missouri}. The cadence of these observations was set by the NEID queue in semester 2021B. The observations were made in high efficiency mode (owing to the star's brightness) and with exposure times between 18 and 28~minutes. The NEID spectra were processed with the NEID Data Reduction Pipeline (DRP)\footnote{\url{https://neid.ipac.caltech.edu/docs/NEID-DRP/}}. The RVs were extracted via the DRP cross correlation function technique and accessed in the DRP level 2 data products. We list the NEID RVs of Kepler-1513 in Table~\ref{tab:rv_neid}.

\begin{deluxetable}{ccccc}
\tabletypesize{\footnotesize}
\tablecaption{RV Measurements From WIYN-NEID. \label{tab:rv_neid}}
\tablehead{
  \colhead{KIC} &
  \colhead{KOI} & 
  \colhead{Kepler} &
  \colhead{BJD$_{\rm TDB}$} & 
  \colhead{RV (m s$^{-1}$)\tablenotemark{a}}
  }
\startdata
4150804 & 3678 & 1513 & 2459433.92958084 & $-1396.2\pm5.3$ \\
4150804 & 3678 & 1513 & 2459475.80880668 & $-1415.0\pm4.0$ \\
4150804 & 3678 & 1513 & 2459504.59402832 & $-1389.1\pm4.2$ \\
4150804 & 3678 & 1513 & 2459523.62673562 & $-1397.3\pm3.9$ \\
4150804 & 3678 & 1513 & 2459536.64127750 & $-1407.8\pm4.0$ \\
\enddata
\tablenotetext{a}{Unlike the Keck-HIRES and APF-Levy RVs, these values have not been corrected for the systemic velocity offset.}
\end{deluxetable}


\subsection{APF-Levy Spectroscopy} \label{sec:apf}

We collected numerous new RV observations of TOI-2180 using the Levy spectrograph at the 2.4~m Automated Planet Finder (APF) telescope at Lick Observatory \citep{Radovan2010,Radovan2014,Vogt2014}. APF conducts queue observing \citep{Burt2015}, which is useful for orbital characterization of long-period giant planets. The extraction of the RVs and activity indicators from APF-Levy spectra closely follows the procedure described for Keck-HIRES \citep{Fulton2015b}. 

Initial RV characterization of TOI-2180 including acquiring a template spectrum was conducted by \citet{Dalba2022a}. As for the Keck-HIRES observations, the APF-Levy RVs are also processed cumulatively. As such, we reprocessed all of the extant RVs along with the new ones. Table~\ref{tab:rv_toi2180} lists all of the TOI-2180 RVs from Keck-HIRES and APF-Levy. Each entry includes an indicator as to whether that data point was previously published by \citet{Dalba2022a}. 

\begin{deluxetable}{ccccc}
\tabletypesize{\footnotesize}
\tablecaption{RV Measurements of TOI-2180 From Keck-HIRES and APF-Levy. \label{tab:rv_toi2180}}
\tablehead{
  \colhead{BJD$_{\rm TDB}$} & 
  \colhead{RV (m s$^{-1}$)\tablenotemark{a}} &
  \colhead{$S_{\rm HK}$\tablenotemark{b}} &
  \colhead{Tel.} &
  \colhead{New?\tablenotemark{c}}
  }
\startdata
2458888.063868 & $-46.2\pm3.8$ & $0.130\pm0.002$ & APF-Levy & N \\
2458894.911472 & $-59.8\pm4.4$ & $0.148\pm0.002$ & APF-Levy & N \\
2458899.027868 & $-69.0\pm3.8$ & $0.146\pm0.002$ & APF-Levy & N \\
2458906.015644 & $-74.9\pm3.2$ & $0.109\pm0.002$ & APF-Levy & N \\
2458914.011097 & $-70.3\pm3.3$ & $0.151\pm0.002$ & APF-Levy & N \\
2458954.765221 & $-93.5\pm3.9$ & $0.147\pm0.002$ & APF-Levy & N \\
2458961.864505 & $-81.3\pm4.2$ & $0.132\pm0.002$ & APF-Levy & N \\
2458964.879814 & $-86.2\pm3.0$ & $0.130\pm0.002$ & APF-Levy & N \\
2458965.811833 & $-81.3\pm4.0$ & $0.136\pm0.002$ & APF-Levy & N \\
2458968.850718 & $-94.0\pm3.5$ & $0.125\pm0.002$ & APF-Levy & N \\
\enddata
\tablenotetext{}{This is a representative subset of the full data set. The full table will be made available in machine readable format.}
\tablenotetext{a}{These values are relative RVs. The systematic velocity offset has been removed for each telescope.}
\tablenotetext{b}{The $S_{\rm HK}$ values from APF and Keck data have different zero-points.}
\tablenotetext{c}{We reprocessed the entire Keck-HIRES and APF-Levy data sets for this star and therefore publish all of the RVs together. However, ``N'' denotes a measurement originally published by \citet{Dalba2022a} while ``Y'' denotes a previously unpublished measurement.}
\end{deluxetable}


\section{Methods} \label{sec:methods}

In this section, we describe the two main modeling tools used to extract information from the photometric and spectroscopic data. In general, \textsf{EXOFASTv2} \citep{Eastman2013,Eastman2017,Eastman2019} is employed when both transit and RV data sets provide for a clear, high S/N detection of the target exoplanet. In these cases, joint modeling of stellar and planetary parameters provide a thorough characterization of the system. In other cases containing non-detections or other special circumstances, we employ \textsf{RadVel} \citep{Fulton2018} to model just the RV data. Specific details and strategies for each system are described in their respective subsections of Section~\ref{sec:res}. 

\subsection{EXOFASTv2 Modeling}\label{sec:exofast}

\textsf{EXOFASTv2} enables the simultaneous modeling of planetary and stellar parameters informed by various sets of photometric and spectroscopic data. For the stellar spectral energy distribution (SED) and isochrone modeling, we make use of this package's ability to access archival photometric data sets from various broadband surveys. We also adopt \textsf{EXOFASTv2}'s default noise floor on stellar parameters, which accounts for systematic variations that have been found \emph{between} different stellar models \citep{Tayar2020}. This prevents unrealistically small uncertainties on stellar (and therefore planetary) parameters. When modeling long cadence Kepler data, we use numerical integration when evaluating the model to avoid smearing out the transit shape.

We prior each \textsf{EXOFASTv2} fit with several pieces of independent information. First, we apply normal priors via the spectroscopically measured values of stellar effective temperature ($T_{\rm eff}$) and iron abundance ([Fe/H]), which we derive from either moderate S/N recon spectra or high S/N template spectra with \textsf{SpecMatch} \citep{Petigura2015}. Then, we prior the parallax with measurements from Gaia early Data Release 3 \citep{Gaia2021}, corrected for bias via the \citet{Lindegren2021} relations. Lastly, we apply extinction priors with the \citet{Schlafly2011} galactic dust maps. Any additional priors are applied on a case-by-case basis.

For \textsf{EXOFASTv2}, convergence is assessed by the number of independent draws of the posterior \citep{Ford2006b} and the Gelman-Rubin statistic \citep{Gelman1992}. The former must be greater than 1,000 and the latter must be less than 1.01 for each parameter.

\subsection{RadVel Modeling}

For systems in which we only achieve an upper limit on mass and RV semi-amplitude or that otherwise require special treatment, we model the RVs with \textsf{RadVel} \citep{Fulton2018}. \textsf{RadVel} models Keplerian signals and linear or quadratic trends in RV time series data. The circumstances of each \textsf{RadVel} model are described for each system. 

Convergence is assessed across four metrics: the Gelman-Rubin statistic, the number of independent draws, the minimum autocorrelation factor, and the maximum relative change in autocorrelation time \citep{Fulton2018}. In all cases, we run our \textsf{RadVel} fits until each of these metrics is satisfied. 

\subsection{Bulk Metallicity Modeling}\label{sec:metals}

For systems in which we uniquely identify a Keplerian signal in the RVs that matches the transit ephemeris and precisely measure planet mass, we model the planetary bulk metallicity ($Z_P \equiv M_z / M_P$, where $M_z$ is the total mass of metals within the planet). Although this value is model dependent, it can provide key insights as to how planets form and evolve \citep[e.g.,][]{Thorngren2016}. Our analysis closely follows that of \citet{Thorngren2019a}. For these planets, we apply 1D planetary structure models assuming an equal parts rock/ice core to follow their thermal evolution. We also assume a homogeneous convective envelope and a radiative atmosphere interpolated from the \citet{Fortney2007} model grid. We draw samples in posterior probability from planet mass, radius, and stellar age and adjust the metal mass to match the measured radius.

The posterior probability distributions from this analysis are presented in the Appendix. The results of this analysis will be discussed below in the Results section for the applicable planets.


\section{Results} \label{sec:res}

In the following subsections, we describe the modeling and parameter estimation for each system. This includes any previously published analysis and any data external to this paper. 


\subsection{Kepler-111 c (KOI-139.01)} 

Kepler-111~c was statistically validated by \citet{Rowe2014} along with an interior, super-Earth size planet. TTVs were identified for Kepler-111~c by \citet{Holczer2016}. Here, we make no attempt to model either the transits or the RV effect of the inner planet. 

\begin{figure*}
    \centering
    \includegraphics[width=\textwidth]{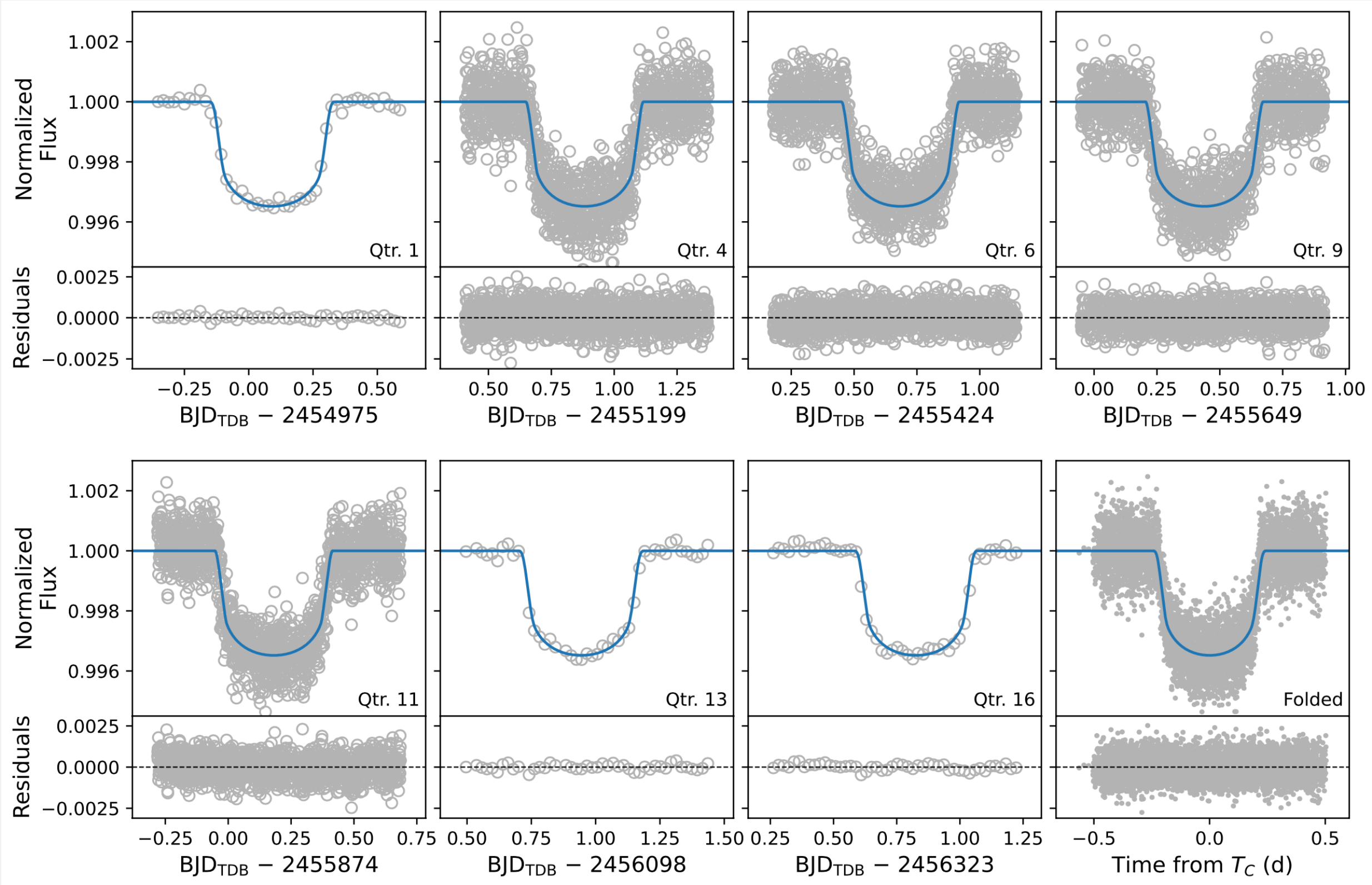}
    \caption{Transits and model of Kepler-111~c, including both short and long cadence Kepler data. Each transit is shown to demonstrate the convergence of the TTV model. In the lower right column, all data have been folded on the maximum likelihood transit ephemeris.}
    \label{fig:transit_CK00139}
\end{figure*}

\begin{figure}
    \centering
    \includegraphics[width=\columnwidth]{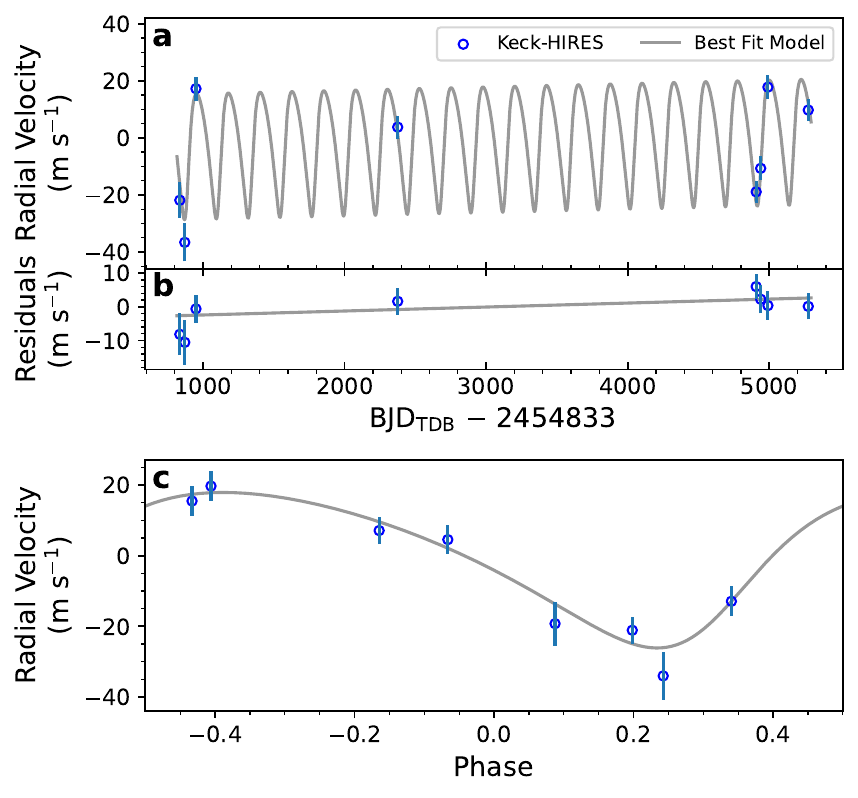}
    \caption{RVs and best fit model for Kepler-111.}
    \label{fig:rv_CK00139}
\end{figure}

The Kepler spacecraft observed seven transits of Kepler-111~c with a mix of short and long cadence data (Figure~\ref{fig:transit_CK00139}). We collected eight RV measurements from Keck-HIRES (Figure~\ref{fig:rv_CK00139}). The Lomb-Scargle periodogram of these RVs is somewhat noisy at periods of a few hundred days owing to the small amount of data collected over a 12-year baseline (Figure~\ref{fig:periodogram}). However the transit signal aligns with the peak power of the periodogram, offering support that we have identified the Keplerian signal of Kepler-111~c.

We conducted the \textsf{EXOFASTv2} fit in two steps. First, we included TTVs and did not apply any additional priors beyond the default set (see Section~\ref{sec:exofast}). Each transit incurs 3 additional parameters: the timing, the out-of-transit flux offset (relative to unity), and a photometric jitter term that increases or decreases the photometric errors. The high dimensionality of this setup drastically slowed overall convergence. We allowed the photometric ``nuisance'' parameters (flux offset and jitter) to converge, and then we stopped the fit. In the second \textsf{EXOFASTv2} fit, we froze those parameters at their converged values. We also noticed that the chains explored unrealistically high values of RV jitter in the first fit. Therefore, we applied a reasonable uniform prior of 0--20~m~s$^{-1}$ on the RV jitter (0--400 m$^2$/s$^2$ on jitter variance) to also assist convergence.

The best fit parameters from the final, converged \textsf{EXOFASTv2} fit are listed in Table~\ref{tab:params_CK00139} in the Appendix. The orbital period and conjunction time published there are the result of a least squares fit to the individual transit times (Table~\ref{tab:ttv_CK00139}). 

We confirm the planetary nature of Kepler-111~c and identify it precisely as a $0.632^{+0.023}_{-0.021}$~\rj\ planet with a mass of $0.70^{+0.14}_{-0.13}$~\mj\ on a $224.77833\pm0.00026$-day orbit. We measure a weak eccentricity but note that its orbit is consistent with circular following the Lucy-Sweeney bias \citep{Lucy1971}. This planet's equilibrium temperature is approximately 352~K. We find no evidence for a long-term trend in the RV residuals.

We model the bulk metallicity of Kepler-111~c using the best fit parameters from the \textsf{EXOFASTv2} fit. We infer a value of $Z_P = 0.60\pm0.04$, meaning that 60\% of this planet's mass is in heavy elements. This planet stands out among other giant planets with mass greater than that of Saturn and perhaps points to a unusual formation or migration scenario.

\subsubsection{The Significance of the Kepler-111~c TTVs}

\begin{figure}
    \centering
    \includegraphics[width=\columnwidth]{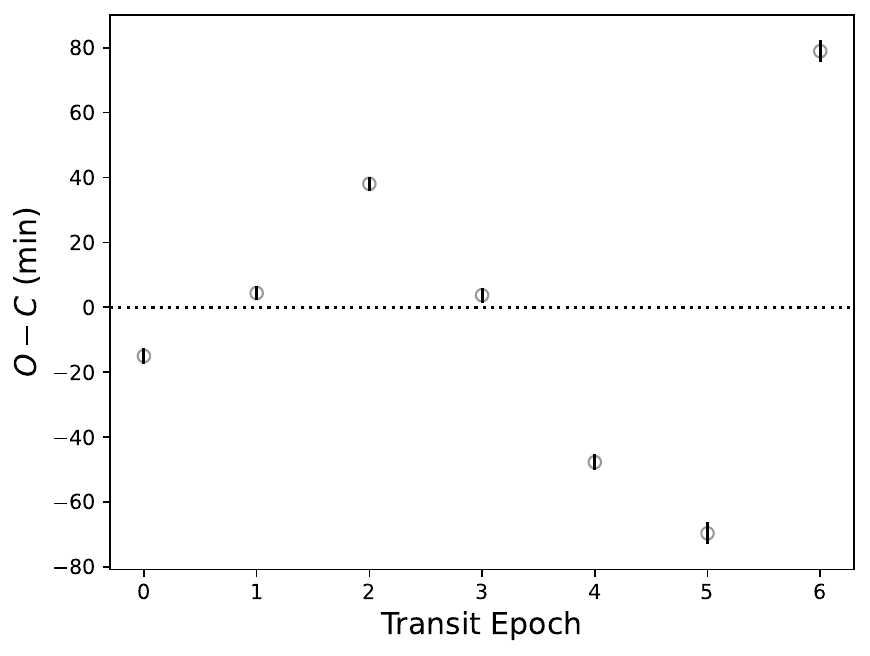}
    \caption{Transit timing variations for Kepler-111~c.}
    \label{fig:ttv_CK00139}
\end{figure}

The final \textsf{EXOFASTv2} fit identified strong TTVs for Kepler-111~c, which we display in Figure~\ref{fig:ttv_CK00139}. Hereafter, $O-C$ refers to the observed transit times relative to the corresponding times calculated with the ephemeris listed in the table of parameters in the Appendix. The seven Kepler transit observations seem to sample one full superperiod of the TTVs. 

By a first order estimate, the inner super-Earth sized planet in the system is unlikely to be the cause of these sizable TTVs. According to \citet[][Equation~10]{Agol2005}, Kepler-111~b would need to be more massive than $\sim$10~$M_{\rm J}$ in order to produce them, which is infeasible. Therefore, another yet undetected object is likely present in this system. Of a variety of possibilities is that of an additional giant planet in mean motion resonance with Kepler-111~c. Within this scenario, the sparsely sampled RVs could instead be picking up on an alias of the additional giant planet, which might explain why Kepler-111~c's mass (and bulk metallicity) was found to be unusually high. Additional RV and transit data would prove helpful in firmly characterizing this entire system.

\begin{deluxetable}{ccc}
\tablecaption{Transit Timing Variations of Kepler-111~c. \label{tab:ttv_CK00139}}
\tablehead{
    \colhead{Epoch} & 
    \colhead{Transit Time (BJD$_{\rm TDB}$)} &
    \colhead{$O - C$ (min)}
    }
\startdata
0 & $2454975.0857 \pm 0.0014$ & $-15.0\pm2.4$ \\
1 & $2455199.8775 \pm 0.0012$ & $4.4\pm2.2$ \\
2 & $2455424.6792 \pm 0.0011$ & $38.1\pm2.2$ \\
3 & $2455649.4337 \pm 0.0011$ & $3.7\pm2.3$ \\
4 & $2455874.1763 \pm 0.0011$ & $-47.7\pm2.5$ \\
5 & $2456098.9394^{+0.0016}_{-0.0017}$ & $-69.6\pm3.3$ \\
6 & $2456323.8209 \pm 0.0015$ & $78.9\pm3.4$ \\
\enddata
\end{deluxetable}


\subsection{Kepler-553 c (KOI-433.02)} 

Kepler-553~c was statistically validated by \citet{Morton2016} along with an interior, Neptune size planet. We included both planets in our \textsf{EXOFASTv2} fit. 

\begin{figure}
    \centering
    \includegraphics[width=\columnwidth]{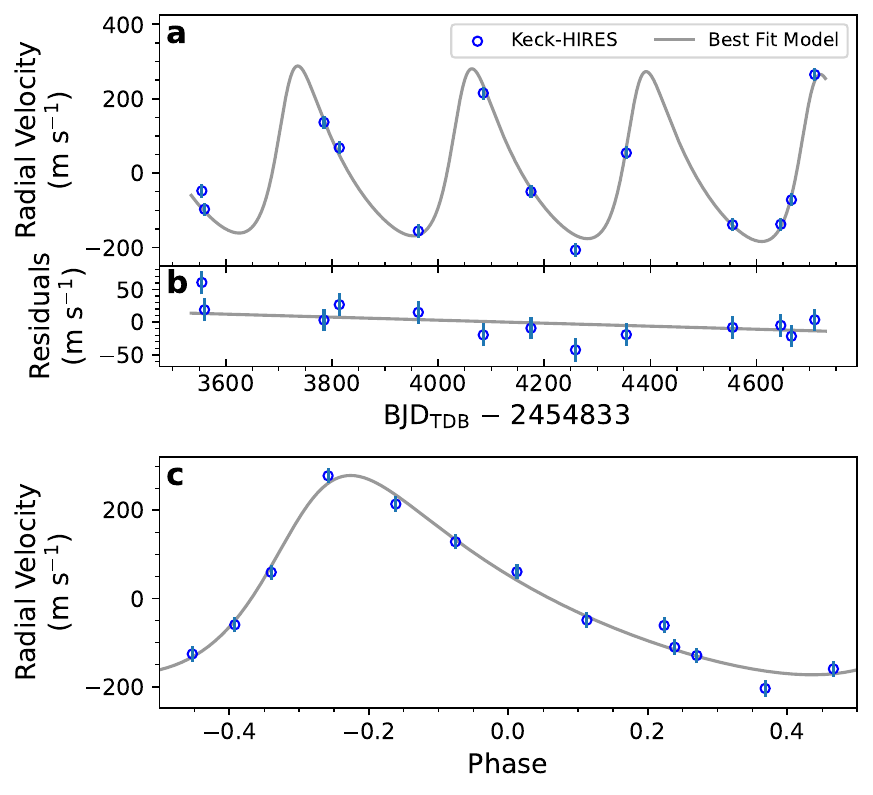}
    \caption{RVs and best fit model for Kepler-553. Kepler-553~b was included in the fit, but no Keplerian signal was detected.}
    \label{fig:rv_K00433}
\end{figure}

\begin{figure*}
    \centering
    \includegraphics[width=0.9\textwidth]{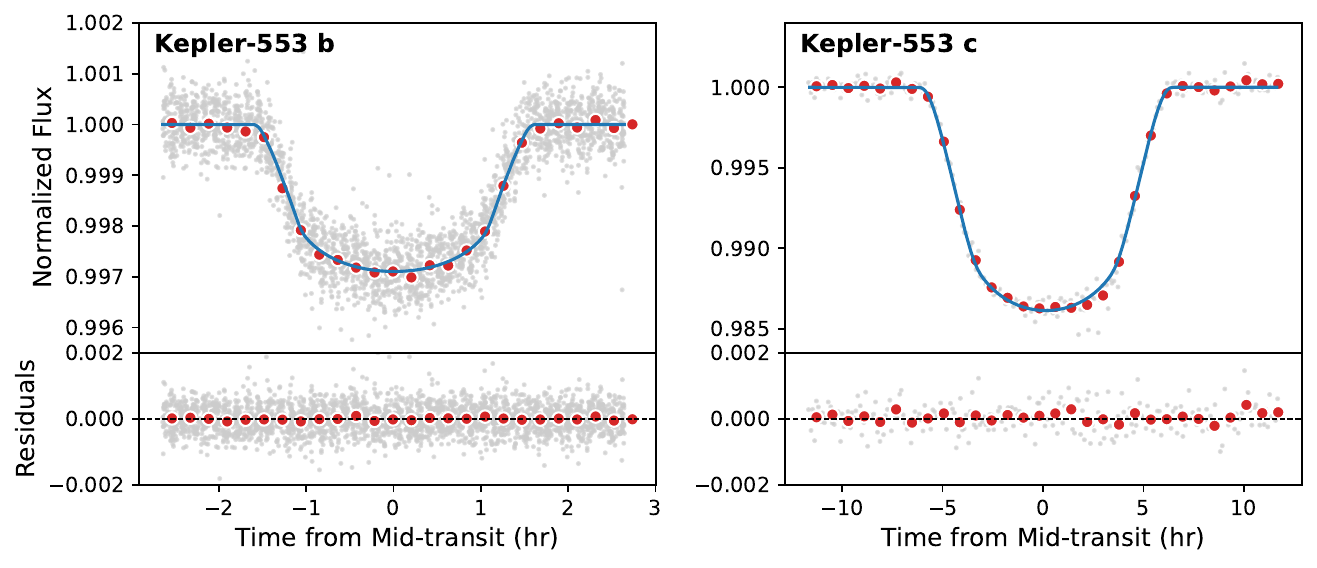}
    \caption{Transits and models of Kepler-533~b and c folded on the maximum likelihood transit ephemeris of each planet. All of the Kepler photometry is long cadence.}
    \label{fig:transit_K00433}
\end{figure*}

The Kepler spacecraft observed five transits of Kepler-553~c with only long cadence data (Figure~\ref{fig:transit_K00433}). We collected 13 RV measurements from Keck-HIRES (Figure~\ref{fig:rv_K00433}). The periodicity of the transits aligns with the maximum power of the Lomb-Scargle periodogram of these RVs (Figure~\ref{fig:periodogram}). The best fit parameters from the final, converged \textsf{EXOFASTv2} fit are listed in Table~\ref{tab:params_K00433} in the Appendix. This fit failed to make a precise measurement for the RV semi-amplitude or mass of the inner planet, which is not surprising given the number of RV observations and the anticipated low mass of the planet. Instead, we report the $3\sigma$ upper limit for these parameters and we exclude any values otherwise derived from the RV data. The transit parameters for Kepler-553~b did converge, though, and are included in the table.

We confirm the planetary nature of Kepler-553~c and identify it precisely as a $1.033^{+0.032}_{-0.025}$~\rj\ planet with a mass of $6.70^{+0.44}_{-0.43}$~\mj\ on a $328.24017^{+0.00039}_{-0.00040}$-day orbit. We measure a moderate eccentricity of $0.346^{+0.020}_{-0.024}$. This planet's equilibrium temperature is approximately 251~K, owing to its relatively cool 5200~K host star, placing it firmly within the habitable zone. We find no evidence for a long-term trend in the RV residuals.

We model the bulk metallicity of Kepler-553~c using the best fit parameters from the \textsf{EXOFASTv2} fit. We infer a value of $Z_P = 0.08\pm0.03$, meaning that 8\% of this planet's mass is in heavy elements. This is consistent with expectations that a giant planet of this mass would be dominated by hydrogen and helium.


\subsection{Kepler-421 b (KOI-1274.01)} 

Kepler-421~b was statistically validated by \citet{Kipping2014a}. \citet{Dalba2016} claim the detection of a partial transit that revealed a lack of TTVs in this system. We did not include the ground-based photometric observations of \citet{Dalba2016} as they did not offer improvement over the highly precise Kepler light curves.

\begin{figure}
    \centering
    \includegraphics[width=\columnwidth]{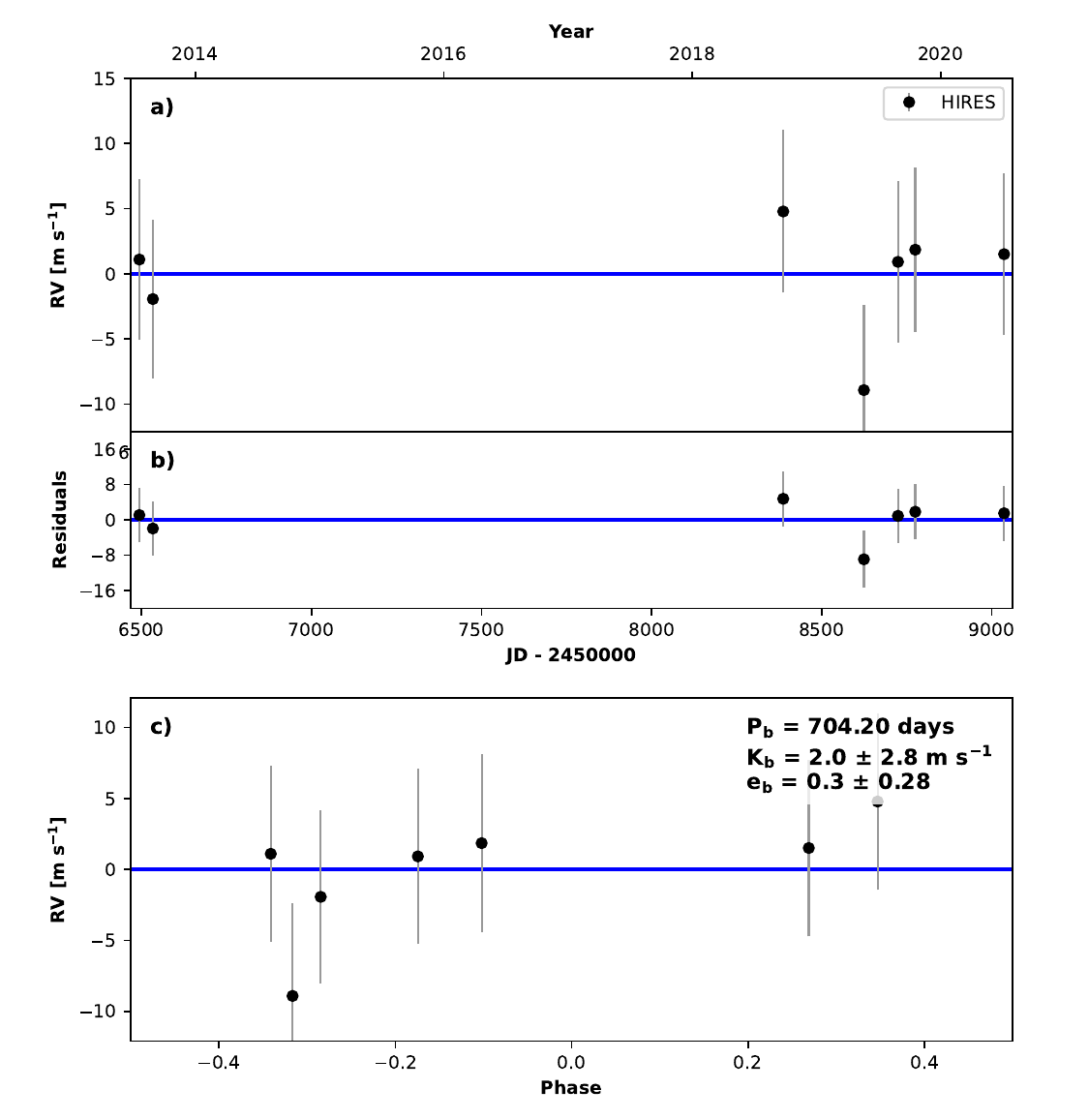}
    \caption{RVs and best fit model for Kepler-421. A unique signal at the ephemeris of the transiting planet is not detected; only an upper limit on planet mass is achieved. The maximum likelihood solution (blue line) has a semi-amplitude consistent with 0~m~s$^{-1}$.}
    \label{fig:rv_K01274}
\end{figure}

The Kepler spacecraft observed two transits of Kepler-421~b with only long cadence data. We collected seven RV measurements from Keck-HIRES (Figure~\ref{fig:rv_K01274}). These RVs were not sufficient to make a confident detection of a Keplerian signal at the ephemeris of the Kepler transits (see the periodogram in Figure~\ref{fig:periodogram}). As a result, we employed \textsf{RadVel} to set an upper limit on the mass of Kepler-421~b. In this fit, we fixed the orbital period and conjunction time at those published by \citet{Kipping2014a}. We provided \textsf{RadVel} with a stellar mass of 0.78$\pm$0.03~$M_{\sun}$, which we found by applying \textsf{SpecMatch} to a moderate S/N recon spectrum acquired with Keck-HIRES. We enforced the Beta distribution prior of \citet{Kipping2013a} on eccentricity to prevent the eccentricity for exploring arbitrarily high and realistic values. Owing to the small number of data points, we did not include jitter in the fit. Instead, we added 5.7~m~s$^{-1}$ to our uncertainties (in quadrature) prior to fit, as recommended by \citet{Dalba2020b} for RVs generated via the matched template technique for this type of star. The fit also did not include a linear trend as one was not identified by inspection. The result of the \textsf{RadVel} fit is shown in Figure~\ref{fig:rv_K01274} and the resulting upper limits are listed in Table~\ref{tab:limits}. 

We fail to dynamically confirm the planetary nature of Kepler-421~b at an orbital period of 704 days. Instead, we find that its mass is less than 136~$M_{\earth}$ at 3$\sigma$.

\begin{deluxetable}{ccccc}
\tablecaption{3$\sigma$ Upper Limits on Mass and RV Semi-amplitude. \label{tab:limits}}
\tablehead{
  \colhead{KIC} &
  \colhead{KOI} & 
  \colhead{Kepler} &
  \colhead{$M_P$ ($M_{\earth}$)} & 
  \colhead{$K$ (m s$^{-1}$)}
  }
\startdata
8800954 & 1274 & 421 & 136 & 14.3 \\
11075279 & 1431 & \nodata & 74 & 8.0 \\
4150804 & 3678 & 1513 & 106 & 16.1 \\
6504954 & 1790 & 952 & 350 & 45.5 \\
\enddata
\end{deluxetable}


\subsection{Kepler-807 b (KOI-1288.01)} 

Kepler-807~b was initially statistically validated by \citet{Morton2016}. More recently, \citet{Canas2023} measured the dynamical mass of Kepler-807~b with RV data from the SOPHIE spectrograph and the APOGEE-N telescope and found it to be a high mass brown dwarf. Here, combine the data from these two facilities with ours from Keck-HIRES to refine the properties of this brown dwarf that was mistakenly validated as an exoplanet. 

\begin{figure}
    \centering
    \includegraphics[width=\columnwidth]{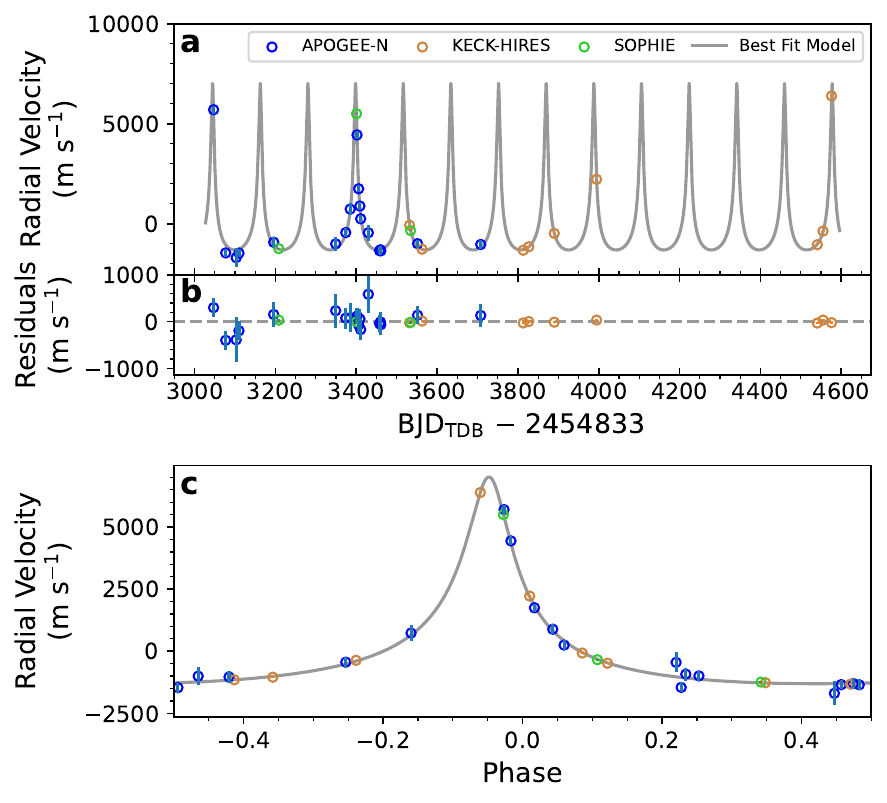}
    \caption{RVs and best fit model for Kepler-807. Errorbars are displayed for the Keck-HIRES and SOPHIE points, but they are too small relative to the axis range to be clearly seen. This Keplerian signal is caused by a brown dwarf, not a planet. The APOGEE-N and SOPHIE data were adopted from \citet{Ehrenreich2011} and \citet{Canas2023}.}
    \label{fig:rv_K01288}
\end{figure}

\begin{figure}
    \centering
    \includegraphics[width=\columnwidth]{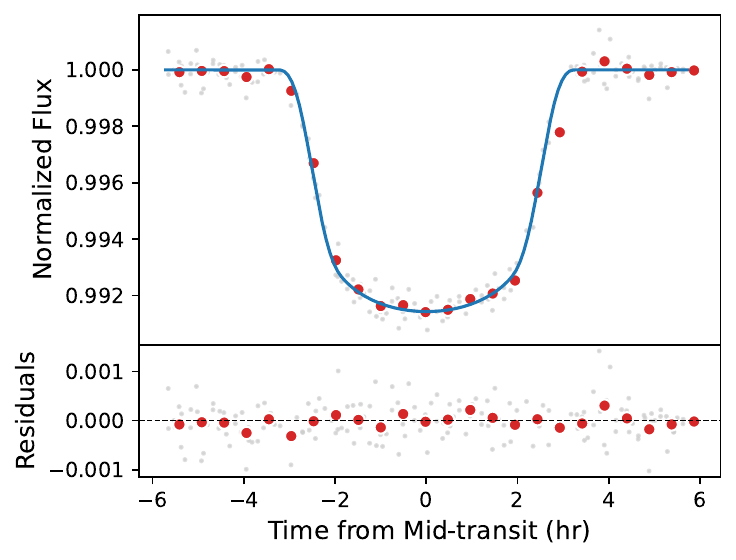}
    \caption{Transits and model of the brown dwarf Kepler-807~b folded on the maximum likelihood transit ephemeris. All of the Kepler photometry is long cadence.}
    \label{fig:transit_K01288}
\end{figure}

The Kepler spacecraft observed eight transits of Kepler-807~b with only long cadence data (Figure~\ref{fig:transit_K01288}). We collected nine RV measurements from Keck-HIRES (Figure~\ref{fig:rv_K01288}). We also adopted  18 RVs from APOGEE-N \citep{Canas2023} and 3 RVs from SOPHIE \citep{Ehrenreich2011}. The Keck-HIRES RVs have higher precision than those from APOGEE-N and SOPHIE by factors of $\sim$35 and $\sim$3, respectively, making them a valuable addition to the extant data. The periodicity of the transits aligns with the maximum power of the Lomb-Scargle periodogram of these RVs (Figure~\ref{fig:periodogram}).

We conducted the \textsf{EXOFASTv2} fit in two steps. First, we did not apply priors for the RV offset ($\gamma$) and jitter for the SOPHIE RVs. Since there are only three RVs from this facility, these parameters unsurprisingly did not converge. In the second \textsf{EXOFASTv2} fit, we applied reasonable uniform  priors for $\gamma$ (5509--7709~m~s$^{-1}$) and jitter (0--500~m~s$^{-1}$ or 0--250,000~m$^2$/s$^2$ for jitter variance) for the SOPHIE RVs based upon visual inspection of the RVs, and convergence was achieved.

The best fit parameters from the final, converged \textsf{EXOFASTv2} fit are listed in Table~\ref{tab:params_K01288} in the Appendix. In agreement with \citet{Canas2023}, we find that Kepler-807~b is actually a brown dwarf of mass $79.8^{+3.4}_{-3.3}$~\mj\ and radius $1.052^{+0.051}_{-0.049}$~\rj. It has a $117.93116\pm0.00020$~day orbit that is highly eccentric ($e = 0.6859^{+0.0020}_{-0.0021}$). 

\subsection{KOI-1431.01} 

KOI-1431.01 is a planet candidate identified by the Kepler science pipeline. It has not been statistically validated. 

\begin{figure}
    \centering
    \includegraphics[width=\columnwidth]{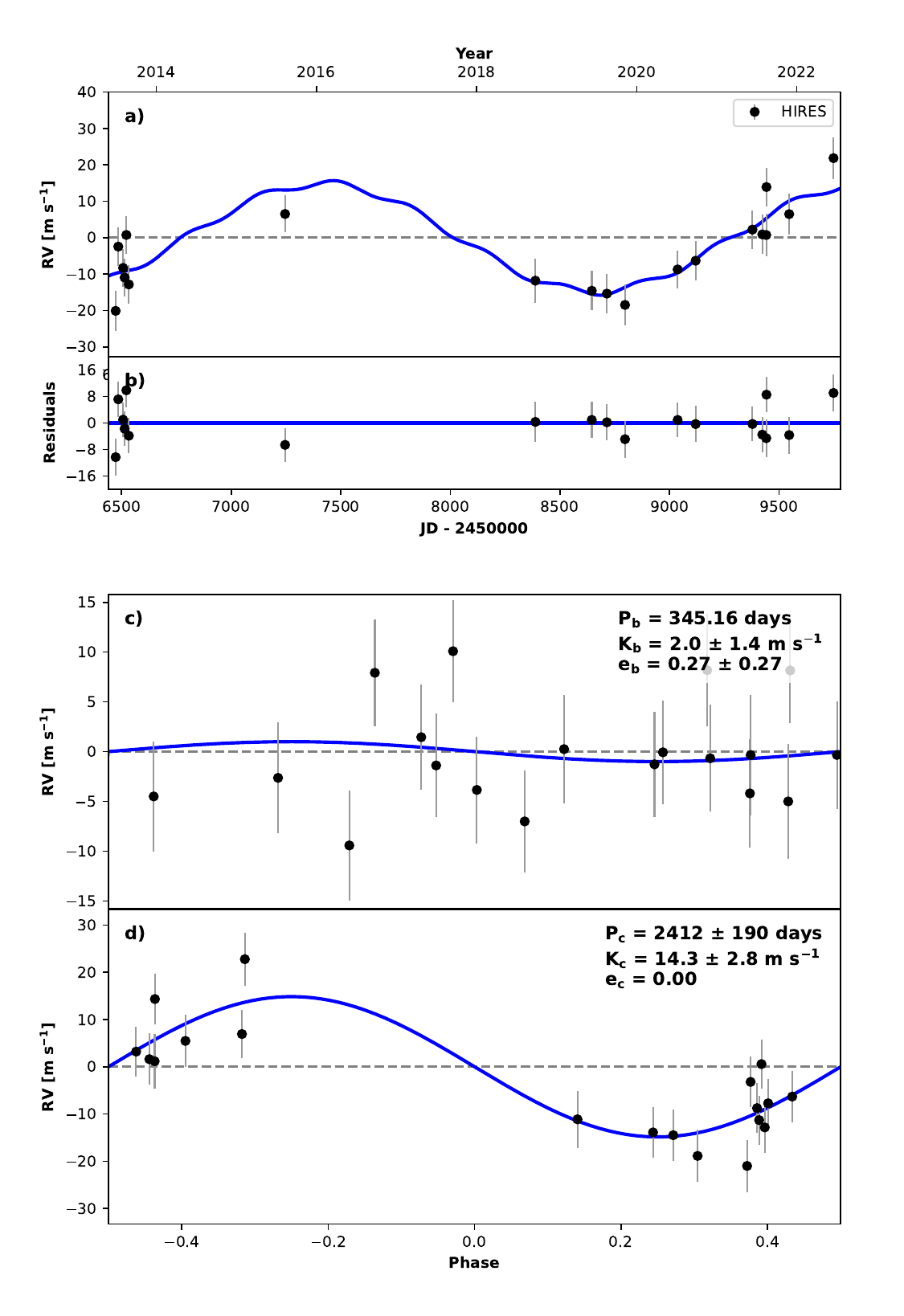}
    \caption{RVs and best fit model for KOI~1431. A unique signal at the ephemeris of the transiting planet is not detected; only an upper limit on planet mass is achieved. The maximum likelihood solution (blue line) of KOI-1431.01 has a semi-amplitude consistent with 0~m~s$^{-1}$ (panel c).} The origin of the long-period signal shown in panel d is unknown (see the text).
    \label{fig:rv_K01431}
\end{figure}

The Kepler spacecraft observed four transits of KOI-1431.01 with only long cadence data. We collected 19 RV measurements from Keck-HIRES (Figure~\ref{fig:rv_K01431}). These RVs were not sufficient to make a confident detection of a Keplerian signal at the ephemeris of the Kepler transits (see the periodogram in Figure~\ref{fig:periodogram}). As a result, we employed \textsf{RadVel} to set an upper limit on the mass of KOI-1431.01. In this fit, we fixed the orbital period and conjunction time at those published in Data Release 25 of the KOI catalog \citep{Thompson2018}. We provided \textsf{RadVel} with a stellar mass of 1.00$\pm$0.05~$M_{\sun}$, which we found by applying \textsf{SpecMatch} to a high S/N template spectrum acquired with Keck-HIRES. We enforced the Beta distribution prior of \citet{Kipping2013a} on eccentricity to prevent the eccentricity for exploring arbitrarily high and realistic values. We allowed for jitter on the HIRES RVs owing to the large number of observations. 

Through visual inspection, a single cycle of a sinusoidal pattern seems to span the entire RV time series. We allow for this signal to be fit by \textsf{RadVel} as a second Keplerian with eccentricity fixed to zero and sampled in log-space for period and semi-amplitude. We hereafter refer to this as the ``long-period signal'' to distinguish it from KOI-1431.01.

The result of the \textsf{RadVel} fit is shown in Figure~\ref{fig:rv_K01431} and the resulting upper limits are listed in Table~\ref{tab:limits}. The long-period signal has an RV semi-amplitude and an orbital period of approximately 15~m~s$^{-1}$ and 2400~days, respectively. If it comes from a bound companion, we find that its minimum mass would be $0.96\pm0.20$~\mj. However, the long-period signal could possibly stem from the host star's magnetic cycle as well. KOI-1431 is known to be rotationally variable. Also, we measured the log~R$^{\prime}_{\rm HK}$ value from its template spectrum to be $-4.83$, which indicates a possibly active star. On the other hand, we do not find correlation between the RVs and chromospheric activity indicators for this system (Figure~\ref{fig:svalue}). Therefore, we cannot make a confident claim as to the origin of the long-period RV signal. In regards to the planet candidate KOI-1431.01, the high eccentricity of the maximum likelihood solution is likely artificial. 

We fail to dynamically confirm the planetary nature of KOI-1431.01 at an orbital period of 345 days. Instead, we find that its mass is less than 74~$M_{\oplus}$ at $3\sigma$. We cannot rule out that this KOI is some form of false positive. 

\subsection{Kepler-849 b (KOI-1439.01)} 

Kepler-849~b was statistically validated by \citet{Morton2016}. TTVs have not been identified in this system.

\begin{figure}
    \centering
    \includegraphics[width=\columnwidth]{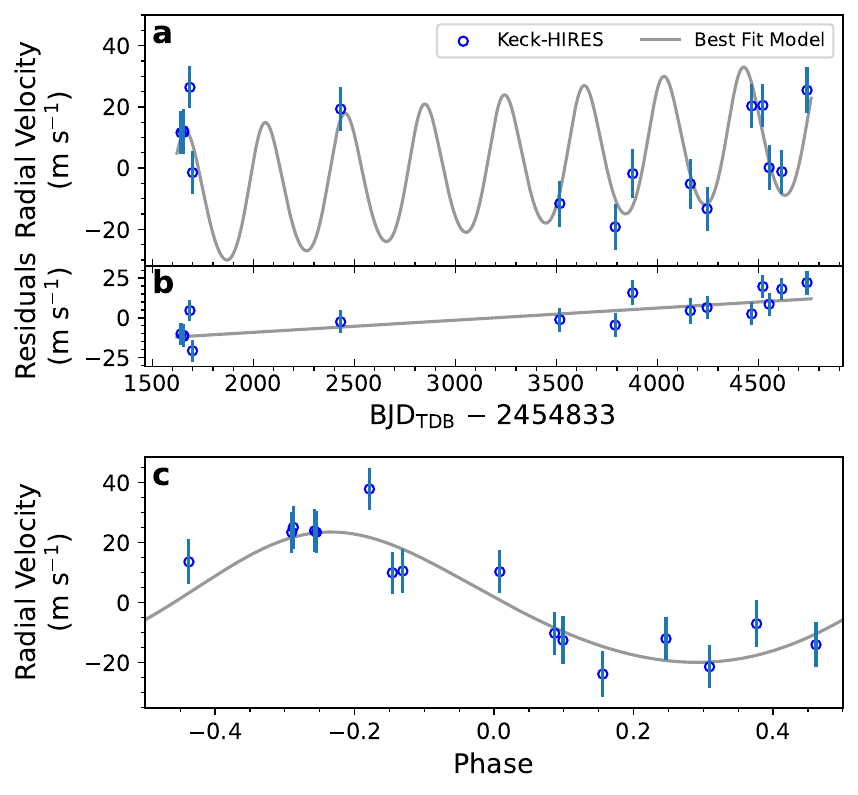}
    \caption{RVs and best fit model for Kepler-849.}
    \label{fig:rv_K01439}
\end{figure}

\begin{figure}
    \centering
    \includegraphics[width=\columnwidth]{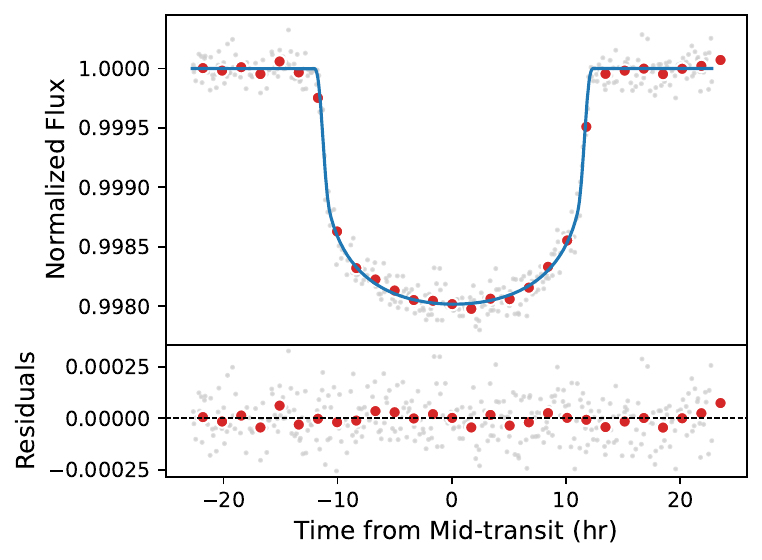}
    \caption{Transits and model of Kepler-849~b folded on the maximum likelihood transit ephemeris. All of the Kepler photometry is long cadence.}
    \label{fig:transit_K01439}
\end{figure}

The Kepler spacecraft observed four transits of Kepler-849~b with only long cadence data (Figure~\ref{fig:transit_K01439}). We collected 16 RV measurements from Keck-HIRES (Figure~\ref{fig:rv_K01439}). The periodicity of the transits aligns with the maximum power of the Lomb-Scargle periodogram of these RVs (Figure~\ref{fig:periodogram}). The best fit parameters from the final, converged \textsf{EXOFASTv2} fit are listed in Table~\ref{tab:params_K01439} in the Appendix. The stellar mass posterior in this converged fit is multimodal. We believe that this multimodality is astrophysical, owing to its slightly evolved evolutionary stage. Therefore, we make no attempt to refine its posterior. We simply note that this is the reason for the slightly larger than expected uncertainties on stellar age, planet mass, planet semi-major axis, and other downstream derived parameters.

We confirm the planetary nature of Kepler-849~b and identify it precisely as a $0.721^{+0.027}_{-0.029}$~\rj\ planet with a mass of $0.94\pm0.20$~\mj\ on a $394.62508^{+0.00081}_{-0.00079}$-day orbit. We find that its orbit is fully consistent with circular. This planet's equilibrium temperature is approximately 363~K. We find evidence for a long-term trend in the RV residuals at 2.6$\sigma$.

We model the bulk metallicity of Kepler-849~b using the best fit parameters from the \textsf{EXOFASTv2} fit. We infer a value of $Z_P = 0.49\pm0.04$, meaning that half of this planet's mass is in heavy elements. Similar to Kepler-111~c, this planet also stands out among other Saturn and Jupiter-mass exoplanets by its small radius (and thereby high bulk metallicity). Future work could consider formation scenarios that might be able to produce such dense gas giant planets.


\subsection{Kepler-952 b (KOI-1790.01)} 

Kepler-952~b was statistically validated by \citet{Morton2016} and found to have TTVs by mutliple studies \citep{Holczer2016,Gajdos2019}.  

\begin{figure}
    \centering
    \includegraphics[width=\columnwidth]{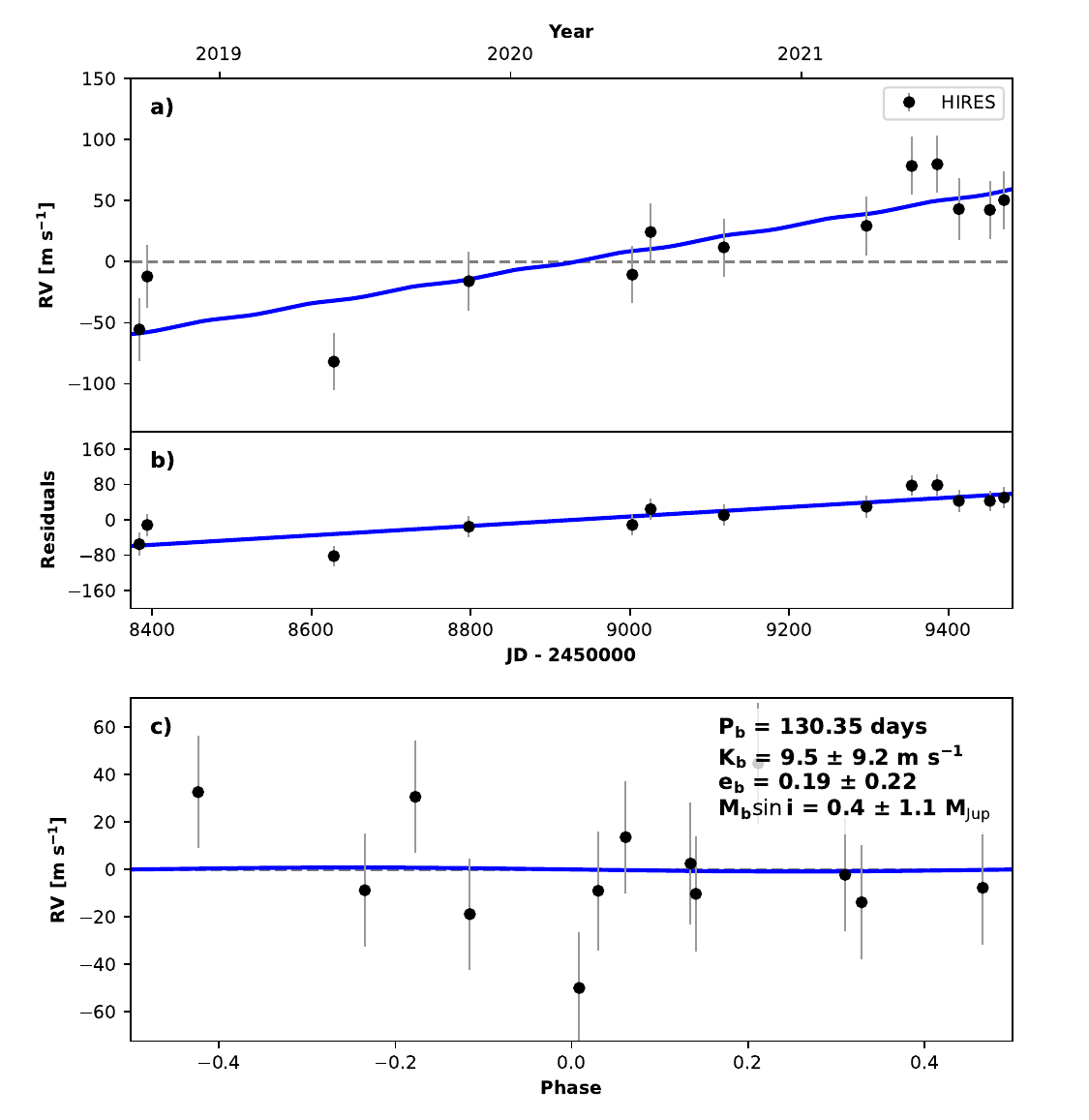}
    \caption{RVs and best fit model for Kepler-952. A unique signal at the ephemeris of the transiting planet is not detected; only an upper limit on planet mass is achieved. The maximum likelihood solution (blue line) has a semi-amplitude consistent with 0~m~s$^{-1}$.}
    \label{fig:rv_K01790}
\end{figure}

The Kepler spacecraft observed eight transits of Kepler-952~b with only long cadence data. We collected 13 RV measurements from Keck-HIRES (Figure~\ref{fig:rv_K01790}). These RVs were not sufficient to make a confident detection of a Keplerian signal at the ephemeris of the Kepler transits (see the periodogram in Figure~\ref{fig:periodogram}). As a result, we employed \textsf{RadVel} to set an upper limit on the mass of Kepler-952~b. In this fit, we fixed the orbital period and conjunction time at those published by \citet{Gajdos2019}. We provided \textsf{RadVel} with a stellar mass of 0.94$\pm$0.04~$M_{\sun}$, which we found by applying \textsf{SpecMatch} to a moderate S/N recon spectrum acquired with Keck-HIRES. We enforced the Beta distribution prior of \citet{Kipping2013a} on eccentricity to prevent the eccentricity for exploring arbitrarily high and realistic values. A long-term trend is clearly present in the RV time series, so we included this term in the \textsf{RadVel} fit. We allowed for jitter to vary in the fit for two reasons. First, there was a sufficient number of data points relative to free parameters. Second, by visual inspection, the true jitter appeared to be far larger than the 6.2~m~s$^{-1}$ recommended by \citet{Dalba2020b} for RVs generated via the matched template technique for this type of star. This is almost certainly due to the faintness of the host star and corresponding low S/N of the spectra.

The result of the \textsf{RadVel} fit is shown in Figure~\ref{fig:rv_K01790} and the resulting upper limits are listed in Table~\ref{tab:limits}. Overall, it likely that the $V=15.3$ host star is simply too faint for the extraction of precise RVs at the $\sim$10's of m~s$^{-1}$ level.

We fail to dynamically confirm the planetary nature of Kepler-952~b at an orbital period of 130 days. Instead, we find that its mass is less than 1.1~$\mj$ at $3\sigma$. We measure a significant long-term trend over the 2.5-year RV baseline of $0.101^{+0.023}_{-0.024}$~m~s$^{-1}$~day$^{-1}$.

\subsection{PH-2 b (KOI-3663.01)} 

PH-2~b (also known as Kepler-86~b) was initially statistically validated by \citet{Wang2013} and later by \citet{Morton2016}. Initial RVs observations from the SOPHIE spectrograph made by \citet{Santerne2016} placed an upper limit on this object's mass. More recently, though, \citet{Dubber2019} published RVs from HARPS-N that yielded a precise dynamical mass in the planetary regime. Here, we combine the HARPS-N RVs with our Keck-HIRES RVs to refine the properties of this giant exoplanet. 

\begin{figure}
    \centering
    \includegraphics[width=\columnwidth]{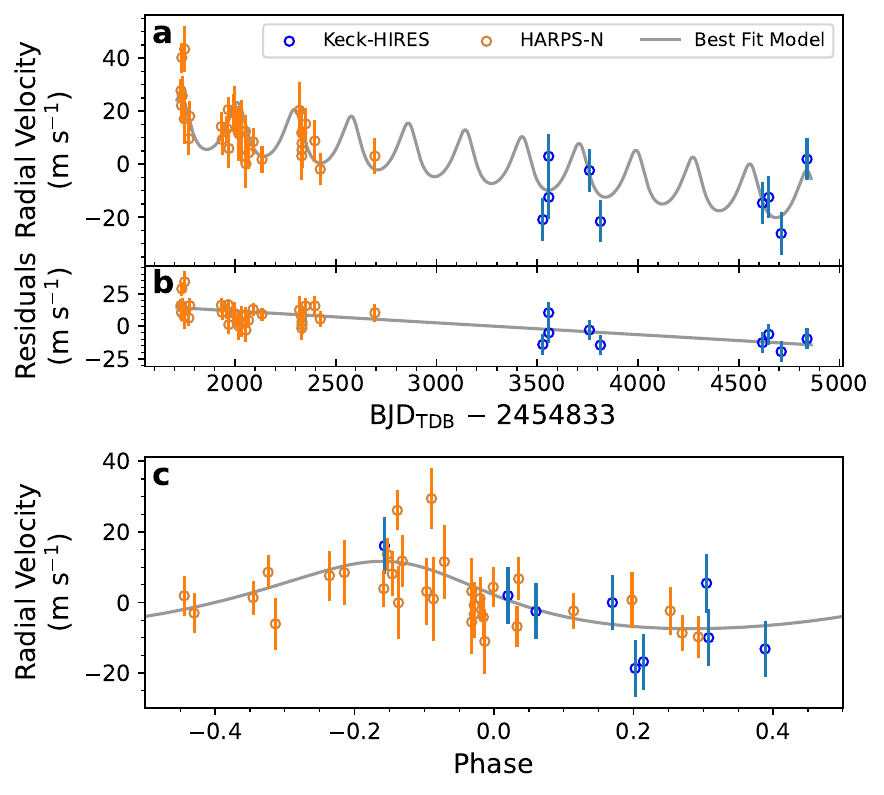}
    \caption{RVs and best fit model for PH-2. The HARPS-N data were adopted from \citet{Dubber2019}.}
    \label{fig:rv_K03663}
\end{figure}

\begin{figure}
    \centering
    \includegraphics[width=\columnwidth]{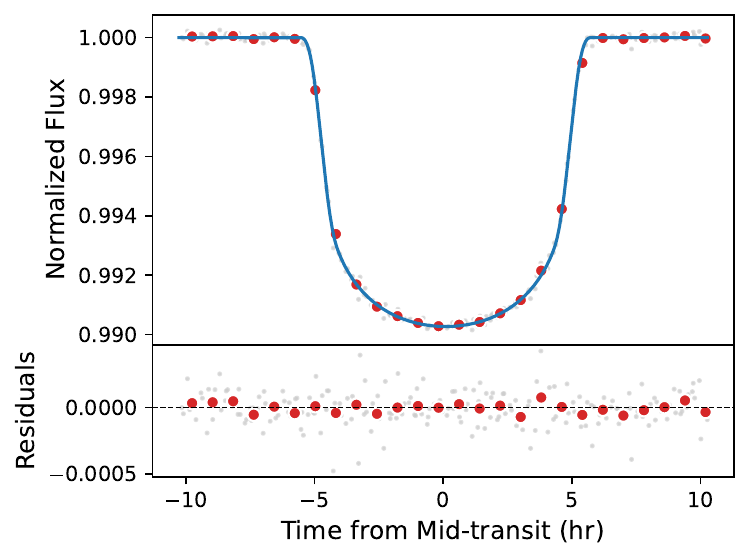}
    \caption{Transits and model of PH-2~b folded on the maximum likelihood transit ephemeris. All of the Kepler photometry is long cadence.}
    \label{fig:transit_K03663}
\end{figure}

The Kepler spacecraft observed five transits of PH-2~b with only long cadence data (Figure~\ref{fig:transit_K03663}). We collected nine RV measurements from Keck-HIRES (Figure~\ref{fig:rv_K03663}). We also adopted  32 RVs from HARPS-N \citep{Dubber2019}. The Keck-HIRES RVs have higher precision than those from HARPS-N and SOPHIE by factors of 2--4, making them a valuable addition to the extant data. The periodicity of the transits aligns with the maximum power of the Lomb-Scargle periodogram of these RVs (Figure~\ref{fig:periodogram}).

We conducted an \textsf{EXOFASTv2} fit that specifically allowed for a long term linear trend in the RV time series. Since there was no temporal overlap between the Keck-HIRES and HARPS-N data sets, this parameter could have been highly unconstrained with the relative RV offsets of either data set. However, the shape of the Keplerlian signal (mostly set by the eccentricity) was sufficiently defined in each data set such that a tentative long term trend was identified at the $2\sigma$ confidence level. Beyond the default priors described in Section~\ref{sec:exofast}, no additional priors were applied in this fit.  

The best fit parameters from the final, converged \textsf{EXOFASTv2} fit are listed in Table~\ref{tab:params_K03663} in the Appendix. Our characterization of this planet is broadly consistent with that of \citet{Dubber2019}. However, we measure a notably lower mass for PH-2~b. We find that PH-2~b has a mass of $0.274^{+0.066}_{-0.067}$~\mj\ and a radius of $0.833^{+0.026}_{-0.024}$~\rj. It has a $282.52542\pm0.00011$~day orbit that is slightly eccentric and that places it within its star's habitable zone ($T_{\rm eq} = 295\pm5$~K). 

We model the bulk metallicity of PH-2~b using the best fit parameters from the \textsf{EXOFASTv2} fit. We infer a value of $Z_P = 0.27\pm0.04$, meaning that 27\% of this planet's mass is in heavy elements. This bulk composition is broadly consistent with other giant planets in this mass range \citep[e.g.,][]{Thorngren2016}.


\subsection{Kepler-1513 b (KOI-3678.01)} 

Kepler-1513~b was statistically validated by \citet{Morton2016} and found to have TTVs by multiple studies \citep{Holczer2016,Gajdos2019}. This planet was initially identified as a possible exomoon host target given the frequency of its TTVs \citep{Kipping2023}, a claim that has since been refuted with additional data \citep{Yahalomi2024}. 

The Kepler spacecraft observed nine transits of Kepler-1513~b with only long cadence data. We collected 34 RV measurements from Keck-HIRES and 5 RV measurements from WIYN-NEID (Figure~\ref{fig:rv_K01790}). These RVs were not sufficient to make a confident detection of a Keplerian signal at the ephemeris of the Kepler transits (see the periodogram in Figure~\ref{fig:periodogram}). As a result, we employed \textsf{RadVel} to set an upper limit on the mass of Kepler-1513~b. In this fit, we fixed the orbital period and conjunction time at those published by \citet{Gajdos2019}. We provided \textsf{RadVel} with a stellar mass of 0.96$\pm$0.04~$M_{\sun}$, which we found by applying \textsf{SpecMatch} to a high S/N template spectrum acquired with Keck-HIRES. We did not fit for a long-term trend because one was not obvious in the data by visual inspection. We allowed for jitter on the HIRES RVs owing to the large number of observations. 

\begin{figure}
    \centering
    \includegraphics[width=\columnwidth]{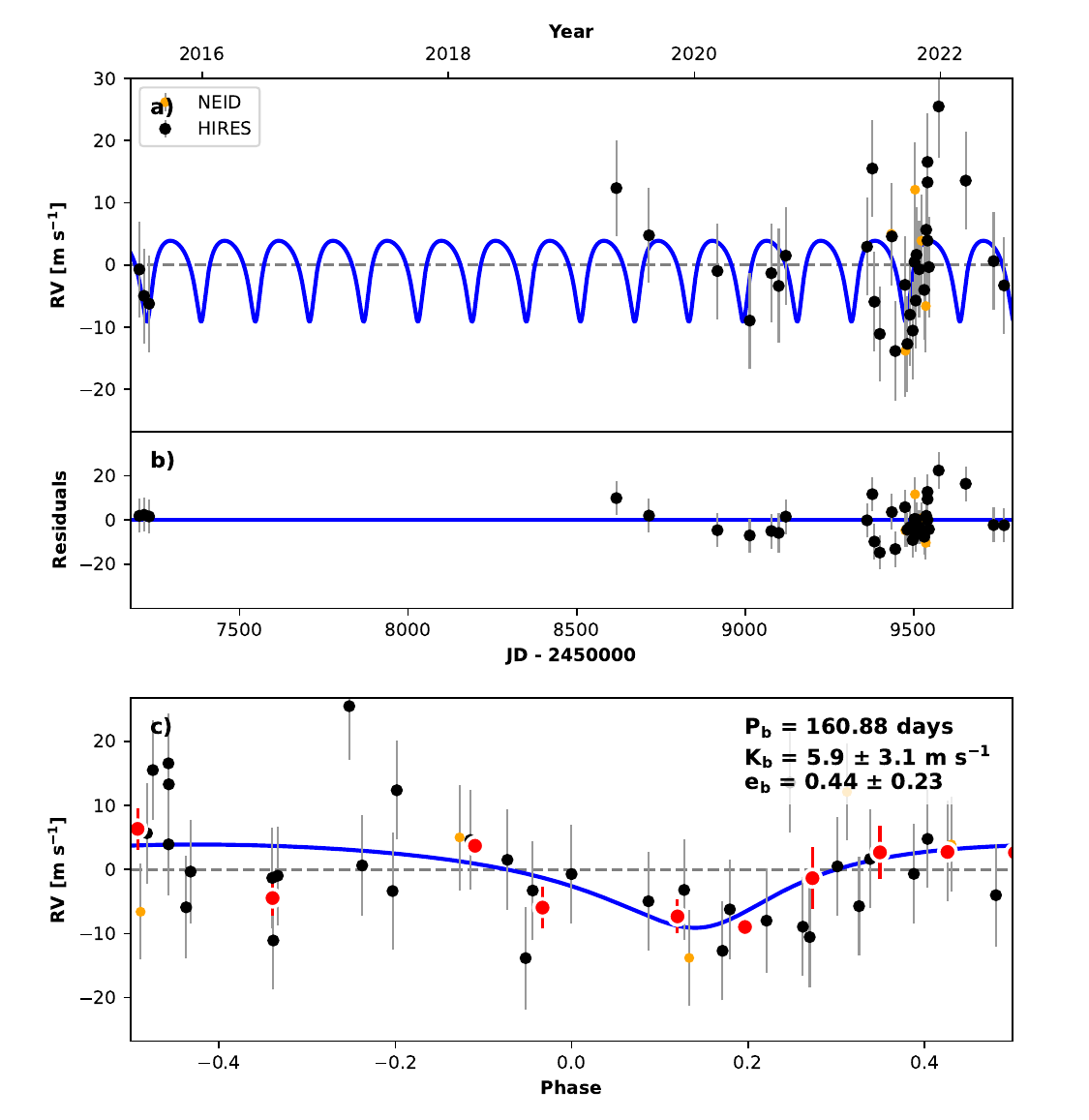}
    \caption{RVs and best fit model for Kepler-1513. The red points are bins of 0.08 in phase. Only a weak signal at the ephemeris of the transiting planet is detected and it relies on a high eccentricity, which may be artificial. Therefore, only an upper limit on planet mass is reported here.}
    \label{fig:rv_CK03678}
\end{figure}

The result of the \textsf{RadVel} fit is shown in Figure~\ref{fig:rv_CK03678} and the resulting upper limits are listed in Table~\ref{tab:limits}. Interestingly, \textsf{RadVel} finds a weak solution with $M_P = 43\pm20$~$M_{\oplus}$ and moderate eccentricity ($e=0.44\pm0.23$). However, the same solution is not found when eccentricity is forced to zero in a subsequent \textsf{RadVel} fit, casting doubt on its robustness. 

We fail to dynamically confirm the planetary nature of Kepler-1513~b at an orbital period of 161 days. Instead, we find that its mass is less than 106~$M_{\oplus}$ at $3\sigma$. This outcome is somewhat surprising given the amount of RV data collected for this system. \emph{A priori}, chromospheric activity may stand as a reasonable explanation for the scatter in the RVs. However, we do not identify correlation between the activity indicators and the RVs in the HIRES spectra (Figure~\ref{fig:svalue}). It is possible, then, that the poor RV data quality stems from the relative low S/N of the spectra and convectively driven stellar jitter (e.g., from granulation). Future work could attempt to model and remove such signals in the RV to potentially locate the subtle Keplerian of the planet.

\subsection{KIC 5951458 b (Kepler-456 b)} 

KIC~5951458~b (also known as Kepler-456~b) was initially statistically validated by \citet{Wang2015b} as a single transit planet. \citet{Dalba2020b} later followed up on this detection with Keck-HIRES and observed a substantial long-term trend in the RVs. This finding cast doubt on the planetary nature of KIC~5951458~b. However, the RV baseline captured by \citet{Dalba2020b} was not sufficient to fully rule out the planetary nature of KIC~5951458~b. They predicted that the 2023 observing season was the optimal time to acquire another RV measurement and revisit the nature of the object creating the single transit event detected by Kepler. 

\begin{figure}
    \centering
    \includegraphics[width=\columnwidth]{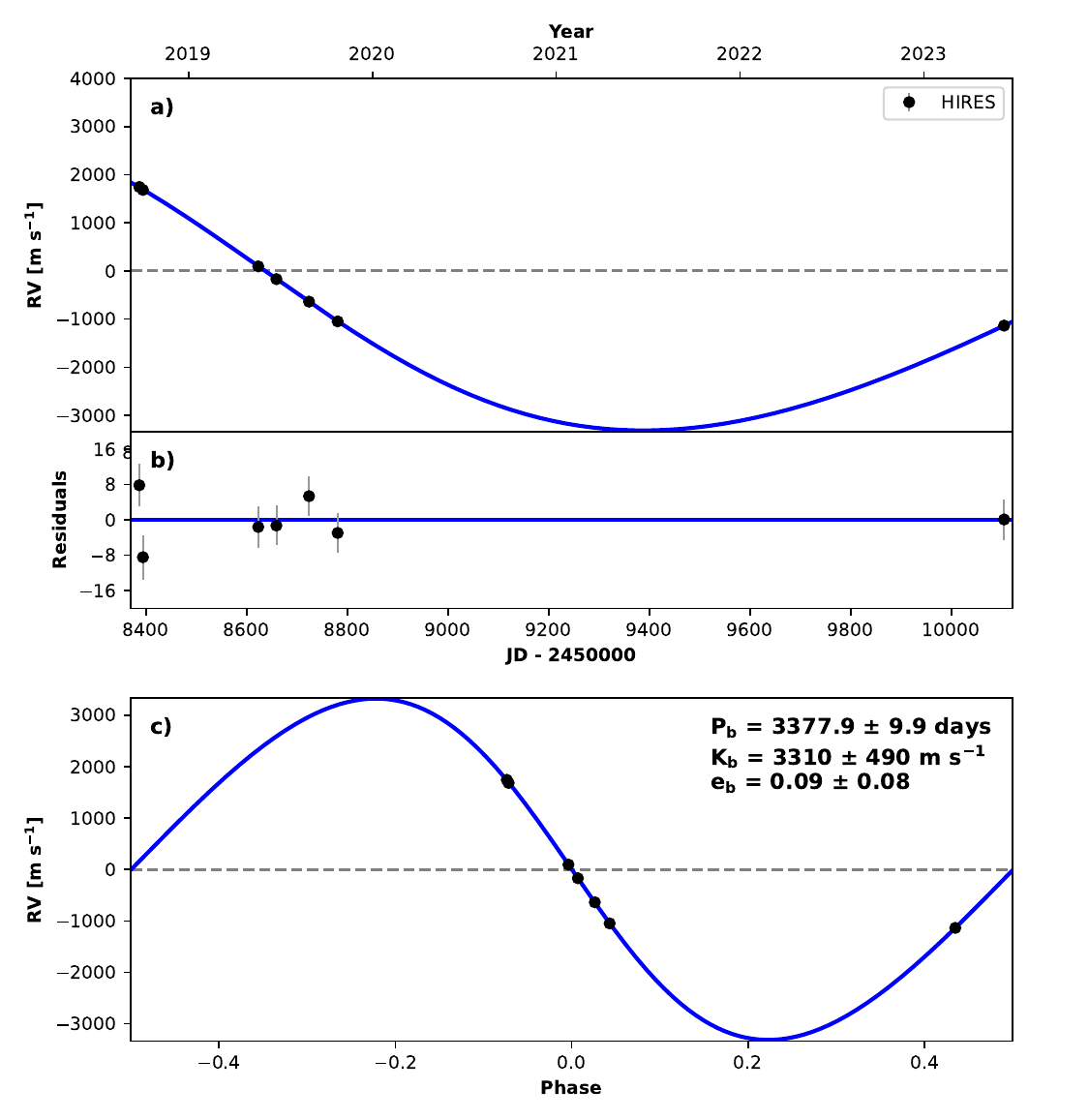}
    \caption{RVs and best fit model for KIC~5951458. Only the final RV is new to this paper. The Keplerian signal is consistent with the conjunction time of the single ``transit'' observed by Kepler. We conclude that KIC~5951458~b is actually an M dwarf companion.}
    \label{fig:rv_KIC595}
\end{figure}

We returned to KIC~5951458~b on 2023 June 27 and collected one additional RV measurement with Keck-HIRES. As described in Section~\ref{sec:hires}, we processed this new RV measurement in concert with all of the previously collected RVs. Since doubt has already been cast on the genuine planetary nature of KIC~5951458~b, we only employed \textsf{RadVel} to explore the RV solution relative to the Kepler transit. In this \textsf{RadVel} fit, we fixed the conjunction time to the value published by \citet{Wang2015b}. Therefore, any solution found for the RVs must be in phase with the Kepler transit. The \textsf{RadVel} fit explored the orbital period and semi-amplitude in log space. We provided \textsf{RadVel} with a stellar mass of 1.16$\pm$0.04~$M_{\sun}$, which we found by applying \textsf{SpecMatch} to a moderate S/N recon spectrum acquired with Keck-HIRES. The \textsf{RadVel} fit readily converged on a circular orbital solution having a mass, RV semi-amplitude, and orbital period of $0.299^{+0.042}_{-0.065}$~$M_{\sun}$, $3310^{+380}_{-600}$~m~s$^{-1}$, and $3378\pm10$~days, respectively (Figure~\ref{fig:rv_KIC595}). This solution is also fully consistent with the low precision Gaia RV measurement presented by \citet{Dalba2020b}. Furthermore, the Gaia DR3 Renormalized Unit Weight Error (RUWE) is 1.4, which indicates a likely binary source \citep{Gaia2022}.

Given this new RV data point and the Gaia RUWE, we argue that KIC~5951458~b is indeed a low mass stellar companion and not a genuine exoplanet. The event observed by Kepler is likely a grazing eclipse. As KIC~5951458~b currently occupies the third spot in the list of longest-period transiting exoplanets\footnote{According to the NASA Exoplanet Archive, accessed 2023 July 26.}, this is an important although unfortunate finding. 

\subsection{TOI-2180 b} 

Although TOI-2180~b is the only exoplanet in this study initially observed by the TESS mission, it is akin to the rest of the long-period Kepler giant planets. TOI-2180~b was initially confirmed and characterized by \citet{Dalba2022a} with RV observations from APF and Keck-HIRES. At that time, only a single transit had been observed by TESS. A subsequent ground-based observational campaign did not make a detection of a second transit. That milestone came later when TESS revisited the system in its extended mission \citep{Dalba2022b}. Here, we refrain from replicating the joint transit, RV, and stellar SED analysis of \citet{Dalba2022a}. Instead, we first present a new analysis of the TESS transit data followed by a new analysis of the RVs.

\subsubsection{TOI-2180~b Transit Analysis}\label{sec:2180_transit}

\begin{figure*}
    \centering
    \includegraphics[width=\textwidth]{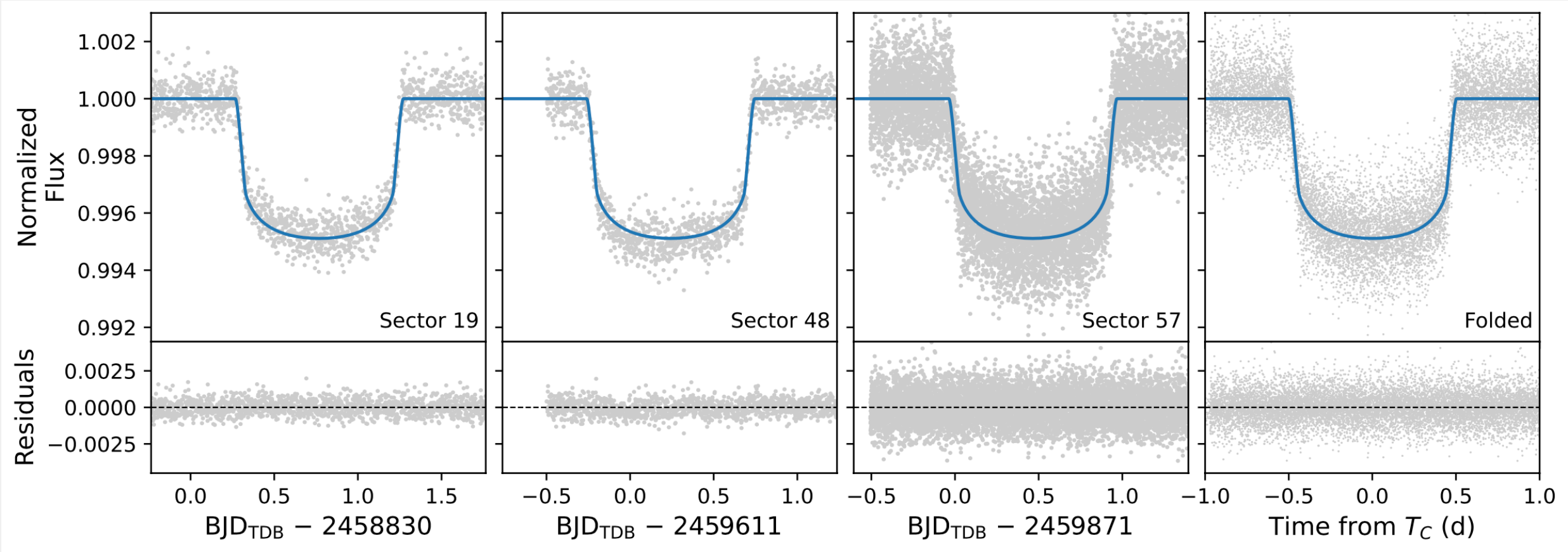}
    \caption{Transits and model of TOI-2180~b, including various cadences as observed by TESS. Each transit is shown to demonstrate the convergence of the TTV model. In the rightmost column, all data have been folded on the maximum likelihood transit ephemeris.}
    \label{fig:transit_T002180}
\end{figure*}

Our treatment of the TESS data is described in Section~\ref{sec:tess}. We inserted these three transits into a joint \textsf{EXOFASTv2} fit that included the transits and stellar SED modeling. The modeling setup mirrored the fit described by \citep{Dalba2022b} and as described in Section~\ref{sec:exofast} with the exception of the RV component. One other critical difference in the fit setup was TTVs. Here, we allow for TTVs in the three transits of TOI-2180~b. The fit converged and we show the resulting transit data and best fit model in Figure~\ref{fig:transit_T002180}. We also list the individual transit times and TTVs in Table~\ref{tab:rv_toi2180}. Our analysis discovers substantial TTVs for TOI-2180~b. 

\begin{figure}
    \centering
    \includegraphics[width=\columnwidth]{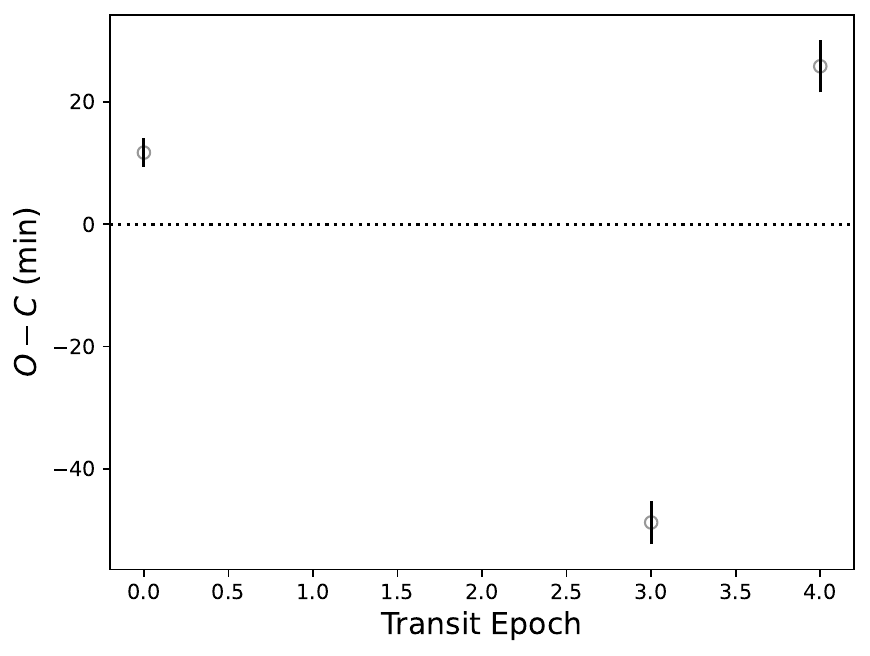}
    \caption{Transit timing variations for TOI-2180~b.}
    \label{fig:ttv_T002180}
\end{figure}

We do not provide a table of stellar or planetary parameters because the values and uncertainties are effectively unchanged from those published by \citet{Dalba2022a}. The only critical update is the orbital ephemeris. We conduct a linear least squares fit to the individual transit times to measure a new orbital period and conjunction time: $260.17060\pm0.00065$~days and $2458830.7561\pm0.0019$~BJD$_{\rm TDB}$, respectively. This ephemeris is used to generate the $O-C$ values displayed in Figure~\ref{fig:ttv_T002180} and listed in Table~\ref{tab:ttv_T002180}.

\begin{deluxetable}{ccc}
\tablecaption{Transit Timing Variations of TOI-2180~b. \label{tab:ttv_T002180}}
\tablehead{
    \colhead{Epoch} & 
    \colhead{Transit Time (BJD$_{\rm TDB}$)} &
    \colhead{$O - C$ (min)}
    }
\startdata
0 & $2458830.7642\pm0.0020$ & $11.7\pm2.3$ \\
3 & $2459611.2370\pm0.0019$ & $-48.8\pm3.6$ \\
4 & $2459871.4604\pm0.0020$ & $25.8\pm4.3$ \\
\enddata
\end{deluxetable}

\subsubsection{TOI-2180~b RV Analysis}\label{sec:2180_rv}

We have continued the Doppler monitoring of TOI-2180 from the APF telescope and the Keck~I telescope in the interim time since the work of \citet{Dalba2022a}. As described in Sections~\ref{sec:hires} and \ref{sec:apf}, the full set of RVs is recalculated with each new measurement. Here, we analyze and present the old (previously published) and new data together.

Given how many RVs we have accumulated for TOI-2180, we applied two quality cuts meant to remove outliers and data of poor quality. We excluded (and do not include in Table~\ref{tab:rv_toi2180}) 12 RVs, all measured by APF-Levy, that occupied the 5th percentile in counts (in the raw spectrum) or the 95th percentile in internal RV error. 

\begin{figure}
    \centering
    \includegraphics[width=\columnwidth]{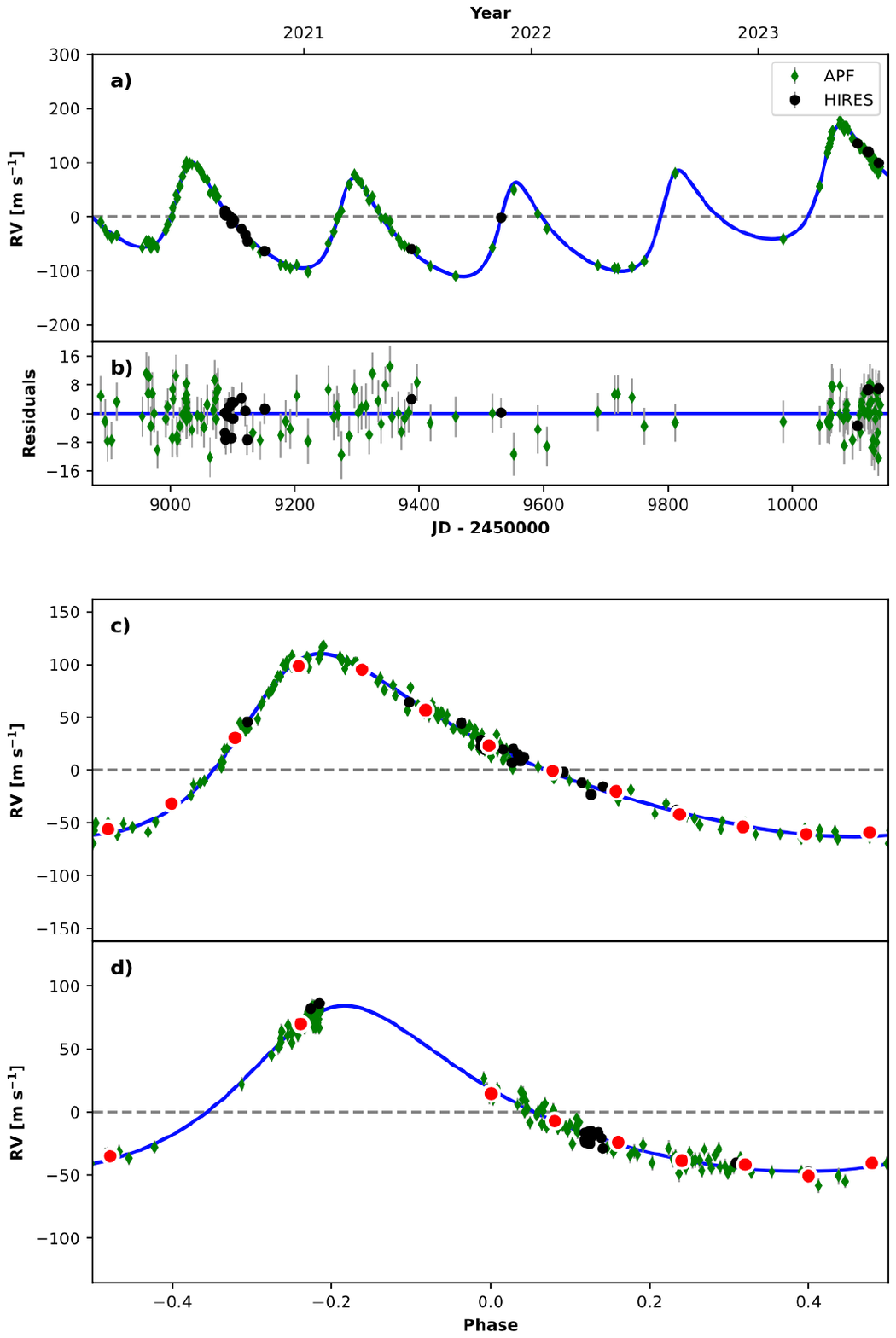}
    \caption{RVs and best fit model for TOI-2180. The Keplerlian signal for TOI-2180~b is shown in panel c. The orbit of the outer companion that was initially identified by \citet{Dalba2022a} is nearly resolved (panel~d). The minimum mass of the outer companion is within the planetary regime.}
    \label{fig:rv_T002180}
\end{figure}

We conducted an RV-only fit to the remaining new and old RVs using \textsf{RadVel}. We applied normal priors to the orbital period and conjunction time using the new ephemeris derived from the three TESS transits (Section~\ref{sec:2180_transit}). We allowed for the fit to model two Keplerian signals: one from TOI-2180~b and one from an outer companion that was previously identified as a long-term trend by \citet{Dalba2022a}. We provided \textsf{RadVel} with a stellar mass of 1.17$\pm$0.06~$M_{\sun}$, which resulted from the stellar and transit joint fit described above. The fit converged without difficulty. The RVs and best fit \textsf{RadVel} model are shown in Figure~\ref{fig:rv_T002180}. The mass of TOI-2180~b derived from this fit is $\sim$2.8~\mj, unchanged from the value published by \citet{Dalba2022a}. However, we almost fully resolve the Keplerian signal from the outer object. We find that this object has an orbital period of $1558^{+68}_{-19}$~days, a conjunction time of $2458904.3^{+6.8}_{-13.0}$~BJD$_{TDB}$, an eccentricity of $0.309^{+0.020}_{-0.029}$, an argument of periastron of $21.5^{+4.6}_{-5.2}$~deg, and an RV semi-amplitude of $65.4^{+2.4}_{-1.7}$~m~s$^{-1}$. We also derive this object's minimum mass as $3.94^{+0.27}_{-0.22}$~\mj. If its mutual inclination with TOI-2180~b is within a few degrees, it is likely to be a giant planet.  

\begin{figure}
    \centering
    \includegraphics[width=\columnwidth]{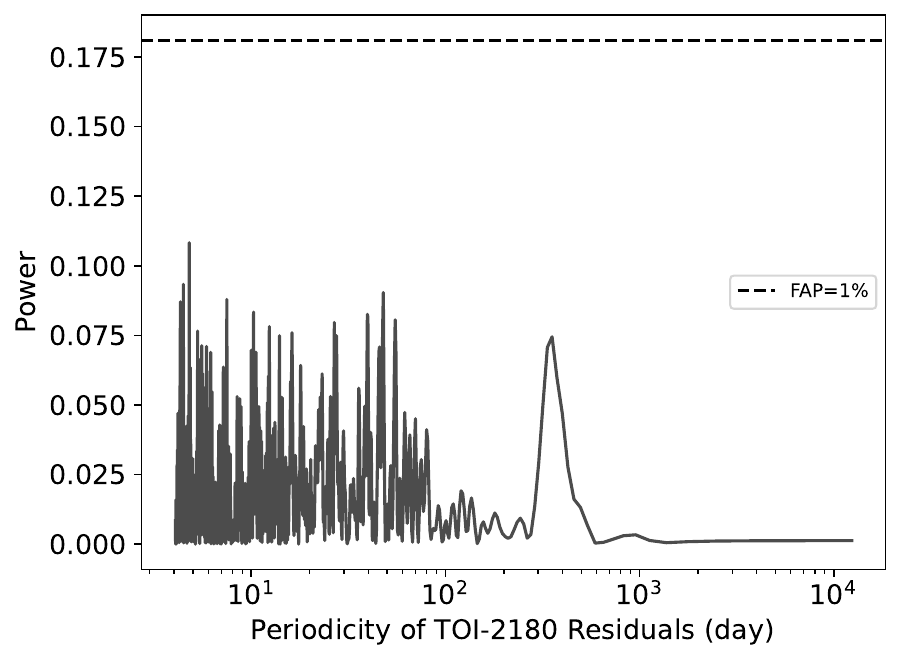}
    \caption{Lomb-Scargle periodogram of the residuals between the TOI-2180 RVs and the best fit model. No significant periodicity is detected. The dashed line denotes a false alarm probability of 1\%.}
    \label{fig:ls_T002180}
\end{figure}

Our last analysis of the RVs of TOI-2180 is a simple search for periodicity beyond the two clearly identified Keplerian signals shown in Figure~\ref{fig:rv_T002180}. The near-nightly cadence of a substantial portion of the RV baseline should allow for the identification of shorter-period signals, at least to within the scatter of the RVs. In Figure~\ref{fig:ls_T002180}, we show a Lomb-Scargle periodogram of the residuals to the two-object \textsf{RadVel} fit. We find no periodicity in the residuals, suggesting that any additional planets interior to $\sim$100~days would have to be sufficiently low mass such that their signals readily hide within the scatter of the RVs. 

\section{Discussion} \label{sec:disc}

The primary goal of this work is to present a basic characterization of several interesting giant exoplanets that transit their host stars. While there is no shortage of known transiting giant planets available to study, these objects all share the unique trait of having long orbital periods relative to the rest of the known population. The orbital period parameter space beyond $\sim$100~days allows for novel studies compared to, for instance, hot Jupiters. However, the first step toward enabling such investigations is measuring precise values for radius, mass, and orbital ephemerides, which is what we have done here where possible. 

\begin{figure*}
    \centering
    \begin{tabular}{cc}
    \includegraphics[width=\columnwidth]{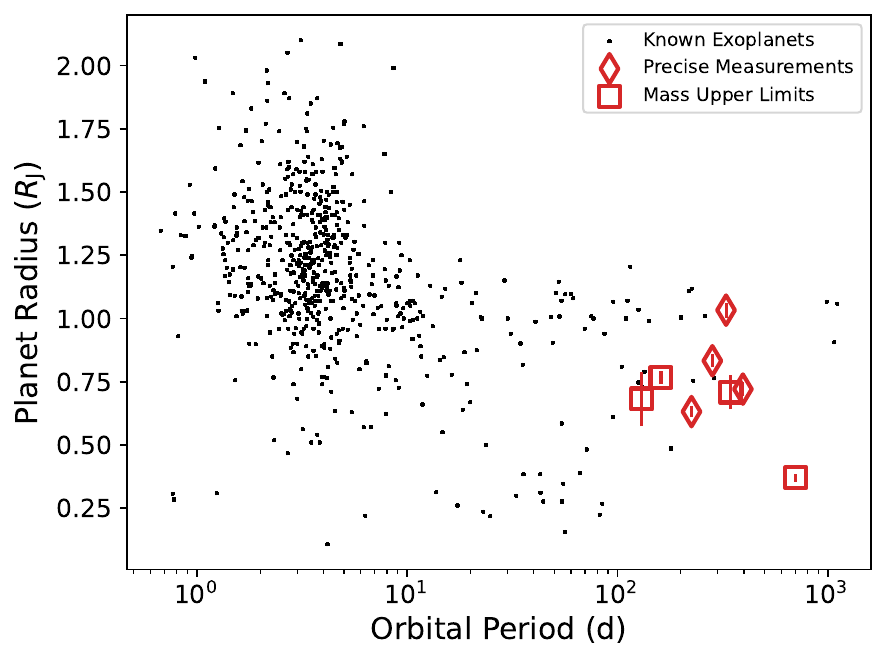} 
    \includegraphics[width=\columnwidth]{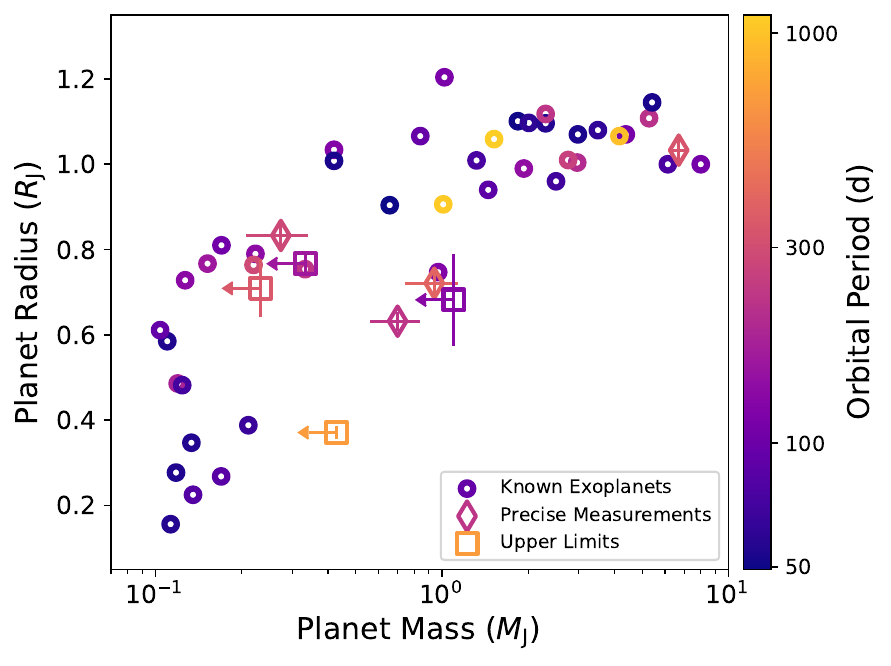}
    \end{tabular}
    \caption{{\it Left:} The results of our work (red shapes) relative to non-controversial planets listed on the NASA Exoplanet Archive (accessed 2023 November 26) with precisely measured radii, orbital periods, and masses greater than 0.1~$\mj$. The red points occupy a scarcely populated region at long orbital periods. {\it Right:} The results of our work (red shapes) relative to non-controversial planets listed on the NASA Exoplanet Archive (accessed 2023 November 26) with precisely measured radii and masses, and with orbital periods greater than 50~days. Masses have been restricted to greater than 0.1~$\mj$. Most of the planets characterized here fall in line with the broad trend of increasing radius with mass until degeneracy pressure dominates in the core of the giant planet. However, Kepler-111~c and Kepler-849~b appear to be overly dense and possibly identify a new category of ``heavy metal'' jovian planets.}
    \label{fig:context}
\end{figure*}

In Figure~\ref{fig:context}, we display the four planets with precisely measured mass (Kepler-111~c, Kepler-553~c, Kepler-849~b, and PH-2~b) and the four planets with mass upper limits (Kepler-421~b, KOI-1431.01, Kepler-1513~b, and Kepler-952~b) relative to other known planetary systems to place these measurements in a broader exoplanetary context. By orbital period (the left panel of Figure~\ref{fig:context}), all of the planets explored by this work clearly stand out among the longest-period planets with known radii. When the actual mass value is considered (the right panel of Figure~\ref{fig:context}), we see that most of the systems fall in line with other cool giant planets. However, Kepler-111~c and Kepler-952~b appear to form a small group of surprisingly dense giant planets. They are joined by Kepler-539~b, which is also has a long, 126~day, orbital period \citep{Mancini2016}. These three planets might be highlighting a particular formation mechanism that results in intense heavy metal accretion. Given their $>$100~day orbital periods, these ``heavy metal jovians'' might exclusively be warm or cool. More efforts like the GOT `EM survey should undertake the challenge of characterizing warm and cool planets so that mysteries like this can be unravelled.

Orbital eccentricity is an especially valuable property to measure for the giant planets in this sample. As the orbital separations are generally far too wide for tidal forces to damp down high eccentricity, this characteristic of the orbit can possibly serve as a tracer of the planets' migration scenarios \citep[e.g.,][]{Dawson2018}. While we detect eccentricities that are convincingly non-zero for Kepler-553~c, Kepler-807~b, PH-2~b, and the outer companion in the TOI-2180 system. All of the planets that are not detected in our survey (i.e., Kepler-421~b, KOI-1431.01, Kepler-952~b, and Kepler-1513~b) could have a range of masses and eccentricities. The eccentricities of planets are simply unknown based on the quality of the data. Future efforts to refine these eccentricities, either through additional RV collection or with complementary methods such as photo-eccentric modeling \citep[e.g.,][]{Ford2008a} would likely be beneficial to our understanding of those systems.

Future modeling efforts could improve our interpretations surrounding the bulk metallicities inferred for Kepler-111~c and Kepler-849~b. For planets of similar mass, these two fall well below the typical radius, hence their high bulk metallicities. Could this be an indication of particularly intense late stage accretion of heavy elements? Or perhaps this requires a fine-tuned formation scenario involving either other objects or well timed disk dissipation? Now that more and more long-period giant planets are having mass and radius measured precisely, theory and modeling work should invest in additional explorations of this unknown parameter space for clues pertaining to giant planet formation.

In this work, we ignore interesting second-order effects from non-spherical planet shapes, rings, or exomoons that may indeed manifest in photometric or spectroscopic data sets. Kepler-1513~b is an interesting case for considering several of these effects. It was identified as a possible exomoon candidate system by \citet{Kipping2023}, although this claim has since been refuted \citep{Yahalomi2024}. A precise planetary mass measurement could be propagated through the exomoon analysis to provide a mass prediction for such a satellite. By its sub-Saturn size, its mass could vary widely \citep[e.g.,][]{Chen2017,Kipping2023}, but it could also appear under-dense if it were to contain a system of rings \citep[e.g.,][]{Piro2020}. Further analysis of the Kepler transit shape or perhaps further transit observations at extremely high precision may reveal more about this planet as a possible host for rings. 

Atmospheric observations are of great interest for warm-to-cool giant exoplanets such as the ones characterized here. Not only does atmospheric characterization of sub-1000~K objects fill a critical gap between the objects in the Solar System and the well characterized hot Jupiter sample, but it could specifically allow for better understanding of giant planet atmospheric physics. \citet{Fortney2020} make testable predictions about CH$_4$ and CO abundances in the atmospheres of cool giant planets ($\lesssim800$~K) that have critical implications for transport in giant planet envelopes. Assessing which, if any, of the giant planets considered here may be amenable to such atmospheric characterization is a valuable extension of this work. However, caution is warranted in this process. While metrics such as the transmission and emission spectroscopy metrics are quite informative \citep{Kempton2017}, they are less useful for comparisons across broad swaths of exoplanet parameter space. The cool giants of the GOT `EM survey will almost always appears as much more costly, difficult targets relative to inner, hotter giant planets. However, at present, there are so few opportunities for advanced characterization of planets on Solar-System-scale orbits, that even a costly target may be an excellent choice \citep[e.g.,][]{Dalba2015}. 

Individually, many of the systems presented here are noteworthy. Beyond Kepler-1513~b, Kepler-111~c and Kepler-553~c both have inner, smaller sibling identified via transits. It may be interesting to consider the formation scenarios for these inner planets given the masses, bulk compositions, and likely formation histories of the outer giant planets \citep[e.g.,][]{Chachan2022}. A similar interest exists for TOI-2180~b, for which we have identified an outer companion, possibly planetary in nature. Future work should explore whether the newly discovered TTVs for TOI-2180~b could be caused by this outer companion or by a so-far undetected inner companion. Otherwise (and perhaps simply owing to its 24~hour transit duration and its host star's brightness), TOI-2180~b stands to become one of the best targets for exomoon searches. 

PH-2~b is one of a small number of well characterized giant exoplanets that exist within their stars' habitable zones \citep[e.g.,][]{Hill2018}. The updated mass measurement presented here is lower than its previously measured value, which likely increases its amenability to atmospheric characterization and also the practical chance it has at receiving in depth follow up characterization. Kepler-553~c is also within its star's habitable zone, with an Earth-like $T_{\rm eq}$ of 251~K. At 6.7~\mj, a massive moon could possibly have formed in this system. Future searches for exomoons should consider this system as a possible target.

Lastly, Kepler-807~b, which is now known to be a brown dwarf, has a mass of $\sim$80~\mj, very near to the canonical hydrogen burning limit. Future work to explore the formation scenario of this object--whether it formed like a giant planet or more so like a stellar object--could provides insights into the different pathways planets and stars take in their own formation.


\section{Summary} \label{sec:summary}

We collected RV observations for 11 planetary systems thought to host transiting giant exoplanets on orbits longer than 100 days. After achieving RV baselines of several years or longer, we were able to dynamically confirm some of these planets and provide upper mass limits on others. We place these measurements in the context of other known exoplanets in Figure~\ref{fig:context}. Here is a case-by-case summary of our findings:

\begin{itemize}
    \item \textit{Kepler-111~c (KOI-139.02):} We confirm the planetary nature of Kepler-111~c and identify it precisely as a 0.6~\rj\ planet with a mass of 0.7~\mj\ on a 225-day orbit. We measure a weak eccentricity but note that its orbit is consistent with circular following the Lucy-Sweeney bias. This planet's equilibrium temperature is approximately 352~K. We find no evidence for a long-term trend in the RV residuals. However, the TTVs of Kepler-111~c point to another undetected object in the system and leave room for wider interpretation of the RV data. We infer a high bulk metallicity of $\sim$60\% for this planet.
    \item \textit{Kepler-553~c (KOI-433.02):} We confirm the planetary nature of Kepler-553~c and identify it precisely as a 1.0~\rj\ planet with a mass of 6.7~\mj\ on a 328-day orbit. We measure a moderate eccentricity of 0.35. This planet's equilibrium temperature is approximately 251~K, owing to its relatively cool 5200~K host star, placing it within its star's habitable zone. We find no evidence for a long-term trend in the RV residuals. We infer a bulk metallicity of $\sim$8\% for this planet.
    \item \textit{Kepler-421~b (KOI-1274.01):} We fail to dynamically confirm the planetary nature of Kepler-421~b at an orbital period of 704 days. Instead, we find that its mass is less than 136~$M_{\earth}$ at $3\sigma$.
    \item \textit{Kepler-807~b (KOI-1288.01):} In agreement with \citet{Canas2023}, we find that Kepler-807~b is actually a brown dwarf of mass 80~\mj\ and radius 1.1~\rj. It has a 118~day orbit that is highly eccentric ($e\approx0.69$). 
    \item \textit{KOI-1431.01:} We fail to dynamically confirm the planetary nature of KOI-1431.01 at an orbital period of 345 days. Instead, we find that its mass is less than 74~$M_{\oplus}$ at $3\sigma$. We also identify a $\sim$2400-day signal in the RVs of unknown origin. We cannot rule out that this KOI is some form of false positive. 
    \item \textit{Kepler-849~b (KOI-1439.01):} We confirm the planetary nature of Kepler-849~b and identify it precisely as a 0.7~\rj\ planet with a mass of 0.9~\mj\ on a 395-day orbit. We find that its orbit is fully consistent with circular. This planet's equilibrium temperature is approximately 363~K. We find evidence for a long-term trend in the RV residuals at 2.6$\sigma$. We infer a high bulk metallicity of $\sim$49\% for this planet.
    \item \textit{Kepler-952~b (KOI-1790.01):} We fail to dynamically confirm the planetary nature of Kepler-952~b at an orbital period of 130 days. Instead, we find that its mass is less than 1.1~$\mj$ at $3\sigma$. We measure a significant long-term trend over the 2.5-year RV baseline of $0.101^{+0.023}_{-0.024}$~m~s~$^{-1}$~day$^{-1}$.
    \item \textit{PH-2~b (KOI-3663.01):} Our characterization of this planet is broadly consistent with that of \citep{Dubber2019} who measured its mass with HARPS-N RVs. However, we measure a notably lower mass for PH-2~b owing to our discovery of a long-term trend in the RV time series. We find the PH-2~b has a mass of 0.27~\mj\ and a radius of 0.8~\rj. It has a 283~day orbit that is slightly eccentric and that places it within its star's habitable zone ($T_{\rm eq} = 295$~K). We infer a bulk metallicity of $\sim$27\% for this planet.
    \item \textit{Kepler-1513~b (KOI-3678.01):} We fail to dynamically confirm the planetary nature of Kepler-1513~b at an orbital period of 161 days. This planet has previously been identified as a potential exomoon host candidate \citep{Kipping2023}, although this claim has been refuted \citep{Yahalomi2024}. Instead, we find that its mass is less than 106~$M_{\oplus}$ at $3\sigma$. Overall, the difficulty in identifying a Keplerian solution in this system is likely tied to the chromospheric activity of the host star. Further work could attempt to model and remove activity signals in the RV to potential locate the subtle Keplerian of the planet. 
    \item \textit{KIC~5951458~b (Kepler-456~b):} We follow the analysis of \citet{Dalba2020b} and obtain a single new RV measurement of this system. Given this new data point and the Gaia RUWE of 1.4, we argue that KIC~5951458~b is a low mass star and not a genuine exoplanet. The event observed by Kepler is likely a grazing eclipse. As KIC~5951458~b is thought to be one of the longest-period transiting exoplanets ever discovered, this is an important albeit unfortunate finding. 
    \item \textit{TOI-2180~b:} We process new TESS data for TOI-2180 that reveal substantial TTVs for the known giant planet TOI-2180~b. We then analyze new and previously published RVs that nearly resolve the $\sim$1558~day orbit of an outer companion that may possibly be planetary by mass. We find no other periodicity within 100~days orbital period, although this is limited by the scatter of the RV data. 
\end{itemize}

Overall, we hope that these discoveries will motivate future studies to further understand the nature of the cool giant planets and their influence on their planetary systems.

\clearpage
\appendix
\section{Tables of Planetary \& Stellar Parameters} \label{sec:appendix}


\startlongtable
\begin{deluxetable*}{lcccccccc}
\tabletypesize{\small}
\tablecaption{Median values and 68\% confidence interval for Kepler-111.}
\tablehead{\colhead{~~~Parameter} & \colhead{Units} & \multicolumn{7}{c}{Values}}
\startdata
\smallskip\\\multicolumn{2}{l}{Stellar Parameters:}&\smallskip\\
~~~~$M_*$ &Mass (\msun) &$1.118^{+0.061}_{-0.068}$\\
~~~~$R_*$ &Radius (\rsun) &$1.145^{+0.040}_{-0.037}$\\
~~~~$L_*$ &Luminosity (\lsun) &$1.440^{+0.11}_{-0.089}$\\
~~~~$F_{Bol}$ &Bolometric Flux (cgs) &$0.0000000001085^{+0.0000000000080}_{-0.0000000000065}$\\
~~~~$\rho_*$ &Density (cgs) &$1.05\pm0.12$\\
~~~~$\log{g}$ &Surface gravity (cgs) &$4.369^{+0.036}_{-0.041}$\\
~~~~$T_{\rm eff}$ &Effective Temperature (K) &$5914^{+81}_{-82}$\\
~~~~$[{\rm Fe/H}]$ &Metallicity (dex) &$0.237^{+0.058}_{-0.059}$\\
~~~~$[{\rm Fe/H}]_{0}$ &Initial Metallicity$^{1}$  &$0.232^{+0.056}_{-0.057}$\\
~~~~$Age$ &Age (Gyr) &$3.4^{+3.1}_{-2.2}$\\
~~~~$EEP$ &Equal Evolutionary Phase$^{2}$  &$357^{+43}_{-34}$\\
~~~~$A_V$ &V-band extinction (mag) &$0.133^{+0.089}_{-0.080}$\\
~~~~$\sigma_{SED}$ &SED photometry error scaling  &$0.88^{+0.35}_{-0.21}$\\
~~~~$\varpi$ &Parallax (mas) &$1.535\pm0.014$\\
~~~~$d$ &Distance (pc) &$651.6\pm5.8$\\
~~~~$\dot{\gamma}$ &RV slope$^{3}$ (m/s/day) &$0.0017^{+0.0018}_{-0.0016}$\\
\smallskip\\\multicolumn{2}{l}{Planetary Parameters:}&c\smallskip\\
~~~~$P$ &Period (days) &$224.77833\pm0.00026$\\
~~~~$R_P$ &Radius (\rj) &$0.632^{+0.023}_{-0.021}$\\
~~~~$M_P$ &Mass (\mj) &$0.70^{+0.14}_{-0.13}$\\
~~~~$T_C$ &Time of conjunction (\bjdtdb) &$2454975.09611\pm0.00088$\\
~~~~$a$ &Semi-major axis (AU) &$0.751^{+0.013}_{-0.016}$\\
~~~~$i$ &Inclination (Degrees) &$89.756^{+0.018}_{-0.020}$\\
~~~~$e$ &Eccentricity  &$0.176^{+0.080}_{-0.085}$\\
~~~~$\omega_*$ &Argument of Periastron (Degrees) &$-141^{+30}_{-19}$\\
~~~~$T_{eq}$ &Equilibrium temperature$^{4}$ (K) &$352.2^{+6.4}_{-5.7}$\\
~~~~$\tau_{\rm circ}$ &Tidal circularization timescale (Gyr) &$229000000^{+110000000}_{-82000000}$\\
~~~~$K$ &RV semi-amplitude (m/s) &$22.2^{+4.5}_{-4.1}$\\
~~~~$R_P/R_*$ &Radius of planet in stellar radii  &$0.05680^{+0.00060}_{-0.00065}$\\
~~~~$a/R_*$ &Semi-major axis in stellar radii  &$141.0^{+5.2}_{-5.6}$\\
~~~~$\delta$ &$\left(R_P/R_*\right)^2$  &$0.003226^{+0.000068}_{-0.000074}$\\
~~~~$\delta_{\rm Kepler}$ &Transit depth in Kepler (fraction) &$0.003543^{+0.000030}_{-0.000027}$\\
~~~~$\tau$ &Ingress/egress transit duration (days) &$0.0417\pm0.0041$\\
~~~~$T_{14}$ &Total transit duration (days) &$0.4638^{+0.0035}_{-0.0033}$\\
~~~~$b$ &Transit Impact parameter  &$0.650^{+0.035}_{-0.044}$\\
~~~~$b_S$ &Eclipse impact parameter  &$0.524^{+0.049}_{-0.043}$\\
~~~~$\tau_S$ &Ingress/egress eclipse duration (days) &$0.0302^{+0.0043}_{-0.0033}$\\
~~~~$T_{S,14}$ &Total eclipse duration (days) &$0.415^{+0.029}_{-0.027}$\\
~~~~$\rho_P$ &Density (cgs) &$3.42^{+0.82}_{-0.72}$\\
~~~~$logg_P$ &Surface gravity  &$3.637^{+0.087}_{-0.095}$\\
~~~~$\Theta$ &Safronov Number  &$1.49^{+0.31}_{-0.28}$\\
~~~~$\fave$ &Incident Flux (\fluxcgs) &$0.00337^{+0.00029}_{-0.00025}$\\
~~~~$T_P$ &Time of Periastron (\bjdtdb) &$2454818^{+27}_{-15}$\\
~~~~$T_S$ &Time of eclipse (\bjdtdb) &$2455068^{+15}_{-11}$\\
~~~~$e\cos{\omega_*}$ &  &$-0.132^{+0.11}_{-0.080}$\\
~~~~$e\sin{\omega_*}$ &  &$-0.103^{+0.065}_{-0.059}$\\
\smallskip\\\multicolumn{2}{l}{Wavelength Parameters:}&Kepler\smallskip\\
~~~~$u_{1}$ &linear limb-darkening coeff  &$0.397\pm0.021$\\
~~~~$u_{2}$ &quadratic limb-darkening coeff  &$0.279\pm0.021$\\
\smallskip\\\multicolumn{2}{l}{Telescope Parameters:}&Keck-HIRES\smallskip\\
~~~~$\gamma_{\rm rel}$ &Relative RV Offset$^{3}$ (m/s) &$3.7^{+3.0}_{-3.4}$\\
~~~~$\sigma_J$ &RV Jitter (m/s) &$7.8^{+5.7}_{-3.7}$\\
\enddata
\label{tab:params_CK00139}
\tablenotetext{}{See Table 3 in \citet{Eastman2019} for a detailed description of all parameters}
\tablenotetext{1}{The metallicity of the star at birth}
\tablenotetext{2}{Corresponds to static points in a star's evolutionary history. See \S2 in \citet{Dotter2016}.}
\tablenotetext{3}{Reference epoch = 2457888.539832}
\tablenotetext{4}{Assumes no albedo and perfect redistribution}
\end{deluxetable*}


\newpage
\startlongtable
\begin{deluxetable*}{lccc}
\tabletypesize{\small}
\tablecaption{Median values and 68\% confidence interval for Kepler-553}
\tablehead{\colhead{~~~Parameter} & \colhead{Units} & \multicolumn{2}{c}{Values}}
\startdata
\smallskip\\\multicolumn{2}{l}{Stellar Parameters:}&\smallskip\\
~~~~$M_*$ &Mass (\msun) &$0.889^{+0.046}_{-0.036}$\\
~~~~$R_*$ &Radius (\rsun) &$0.902^{+0.026}_{-0.021}$\\
~~~~$L_*$ &Luminosity (\lsun) &$0.536\pm0.041$\\
~~~~$F_{Bol}$ &Bolometric Flux (cgs) &$0.0000000000322^{+0.0000000000023}_{-0.0000000000024}$\\
~~~~$\rho_*$ &Density (cgs) &$1.727^{+0.097}_{-0.15}$\\
~~~~$\log{g}$ &Surface gravity (cgs) &$4.480^{+0.019}_{-0.027}$\\
~~~~$T_{\rm eff}$ &Effective Temperature (K) &$5191^{+76}_{-78}$\\
~~~~$[{\rm Fe/H}]$ &Metallicity (dex) &$0.152\pm0.058$\\
~~~~$[{\rm Fe/H}]_{0}$ &Initial Metallicity$^{1}$  &$0.166^{+0.059}_{-0.060}$\\
~~~~$Age$ &Age (Gyr) &$8.8^{+3.3}_{-4.0}$\\
~~~~$EEP$ &Equal Evolutionary Phase$^{2}$  &$366^{+22}_{-24}$\\
~~~~$A_V$ &V-band extinction (mag) &$0.30^{+0.11}_{-0.13}$\\
~~~~$\sigma_{SED}$ &SED photometry error scaling  &$2.94^{+1.3}_{-0.75}$\\
~~~~$\varpi$ &Parallax (mas) &$1.369\pm0.022$\\
~~~~$d$ &Distance (pc) &$730\pm12$\\
~~~~$\dot{\gamma}$ &RV slope$^{3}$ (m/s/day) &$-0.029\pm0.029$\\
\smallskip\\\multicolumn{2}{l}{Planetary Parameters:}&b&c\smallskip\\
~~~~$P$ &Period (days) &$4.0304670\pm0.0000018$&$328.24017^{+0.00039}_{-0.00040}$\\
~~~~$R_P$ &Radius (\rj) &$0.423^{+0.016}_{-0.011}$&$1.033^{+0.032}_{-0.025}$\\
~~~~$M_P$ &Mass$^{4}$ (\mj) &$<0.365$&$6.70^{+0.44}_{-0.43}$\\
~~~~$T_C$ &Time of conjunction (\bjdtdb) &$2455004.09212\pm0.00038$&$2455032.1977\pm0.0010$\\
~~~~$a$ &Semi-major axis (AU) &$0.04766^{+0.00081}_{-0.00065}$&$0.898^{+0.015}_{-0.012}$\\
~~~~$i$ &Inclination (Degrees) &$88.94^{+0.67}_{-0.60}$&$89.8314^{+0.0054}_{-0.0092}$\\
~~~~$e$ &Eccentricity  &\nodata&$0.346^{+0.020}_{-0.024}$\\
~~~~$\omega_*$ &Argument of Periastron (Degrees) &\nodata&$-48.2^{+3.6}_{-3.0}$\\
~~~~$T_{eq}$ &Equilibrium temperature$^{5}$ (K) &$1089\pm19$&$251.0\pm4.3$\\
~~~~$\tau_{\rm circ}$ &Tidal circularization timescale (Gyr) &\nodata&$328000000^{+56000000}_{-47000000}$\\
~~~~$K$ &RV semi-amplitude$^{4}$ (m/s) &$<49$&$226\pm13$\\
~~~~$R_P/R_*$ &Radius of planet in stellar radii  &$0.04828^{+0.00051}_{-0.00040}$&$0.11778^{+0.00067}_{-0.00074}$\\
~~~~$a/R_*$ &Semi-major axis in stellar radii  &$11.40^{+0.21}_{-0.33}$&$214.8^{+3.9}_{-6.3}$\\
~~~~$\delta$ &$\left(R_P/R_*\right)^2$  &$0.002331^{+0.000049}_{-0.000038}$&$0.01387^{+0.00016}_{-0.00017}$\\
~~~~$\delta_{\rm Kepler}$ &Transit depth in Kepler (fraction) &$0.003148^{+0.000062}_{-0.000060}$&$0.01412^{+0.00012}_{-0.00011}$\\
~~~~$\tau$ &Ingress/egress transit duration (days) &$0.00557^{+0.00041}_{-0.00024}$&$0.1127^{+0.0045}_{-0.0046}$\\
~~~~$T_{14}$ &Total transit duration (days) &$0.11567^{+0.00071}_{-0.00069}$&$0.5103\pm0.0029$\\
~~~~$b$ &Transit Impact parameter  &$0.21^{+0.11}_{-0.13}$&$0.751^{+0.010}_{-0.012}$\\
~~~~$b_S$ &Eclipse impact parameter  &\nodata&$0.442^{+0.026}_{-0.017}$\\
~~~~$\tau_S$ &Ingress/egress eclipse duration (days) &\nodata&$0.0477^{+0.0036}_{-0.0022}$\\
~~~~$T_{S,14}$ &Total eclipse duration (days) &\nodata&$0.372^{+0.017}_{-0.012}$\\
~~~~$\rho_P$ &Density (cgs) &\nodata&$7.53^{+0.65}_{-0.74}$\\
~~~~$logg_P$ &Surface gravity  &\nodata&$4.191^{+0.030}_{-0.036}$\\
~~~~$\Theta$ &Safronov Number  &\nodata&$13.03^{+0.81}_{-0.83}$\\
~~~~$\fave$ &Incident Flux (\fluxcgs) &$0.320^{+0.022}_{-0.021}$&$0.000802^{+0.000063}_{-0.000057}$\\
~~~~$T_P$ &Time of Periastron (\bjdtdb) &\nodata&$2454935.8^{+4.2}_{-4.0}$\\
~~~~$T_S$ &Time of eclipse (\bjdtdb) &\nodata&$2454917.5^{+3.2}_{-3.4}$\\
~~~~$e\cos{\omega_*}$ &  &\nodata&$0.231^{+0.015}_{-0.016}$\\
~~~~$e\sin{\omega_*}$ &  &\nodata&$-0.258^{+0.028}_{-0.022}$\\
\smallskip\\\multicolumn{2}{l}{Wavelength Parameters:}&Kepler\smallskip\\
~~~~$u_{1}$ &linear limb-darkening coeff  &$0.540\pm0.033$\\
~~~~$u_{2}$ &quadratic limb-darkening coeff  &$0.162^{+0.048}_{-0.047}$\\
\smallskip\\\multicolumn{2}{l}{Telescope Parameters:}&Keck-HIRES\smallskip\\
~~~~$\gamma_{\rm rel}$ &Relative RV Offset$^{3}$ (m/s) &$13.7^{+8.0}_{-8.1}$\\
~~~~$\sigma_J$ &RV Jitter (m/s) &$26.1^{+12}_{-7.5}$\\
\enddata
\label{tab:params_K00433}
\tablenotetext{}{See Table 3 in \citet{Eastman2019} for a detailed description of all parameters}
\tablenotetext{1}{The metallicity of the star at birth}
\tablenotetext{2}{Corresponds to static points in a star's evolutionary history. See \S2 in \citet{Dotter2016}.}
\tablenotetext{3}{Reference epoch = 2458965.348500}
\tablenotetext{4}{Upper limits on mass and semi-amplitude for Kepler-553~b are 3$\sigma$.}
\tablenotetext{5}{Assumes no albedo and perfect redistribution}
\end{deluxetable*}


\newpage
\startlongtable
\begin{deluxetable*}{lcccc}
\tabletypesize{\small}
\tablecaption{Median values and 68\% confidence interval for Kepler-807.}
\tablehead{\colhead{~~~Parameter} & \colhead{Units} & \multicolumn{3}{c}{Values}}
\startdata
\smallskip\\\multicolumn{2}{l}{Stellar Parameters:}&\smallskip\\
~~~~$M_*$ &Mass (\msun) &$1.068^{+0.071}_{-0.068}$\\
~~~~$R_*$ &Radius (\rsun) &$1.201^{+0.046}_{-0.045}$\\
~~~~$L_*$ &Luminosity (\lsun) &$1.68^{+0.13}_{-0.12}$\\
~~~~$F_{Bol}$ &Bolometric Flux (cgs) &$0.0000000000237\pm0.0000000000011$\\
~~~~$\rho_*$ &Density (cgs) &$0.867^{+0.12}_{-0.098}$\\
~~~~$\log{g}$ &Surface gravity (cgs) &$4.306^{+0.041}_{-0.039}$\\
~~~~$T_{\rm eff}$ &Effective Temperature (K) &$5994\pm78$\\
~~~~$[{\rm Fe/H}]$ &Metallicity (dex) &$0.044^{+0.056}_{-0.053}$\\
~~~~$[{\rm Fe/H}]_{0}$ &Initial Metallicity$^{1}$  &$0.085^{+0.054}_{-0.052}$\\
~~~~$Age$ &Age (Gyr) &$5.4^{+2.9}_{-2.5}$\\
~~~~$EEP$ &Equal Evolutionary Phase$^{2}$  &$397^{+23}_{-45}$\\
~~~~$A_V$ &V-band extinction (mag) &$0.091\pm0.057$\\
~~~~$\sigma_{SED}$ &SED photometry error scaling  &$0.47^{+0.23}_{-0.13}$\\
~~~~$\varpi$ &Parallax (mas) &$0.665\pm0.022$\\
~~~~$d$ &Distance (pc) &$1502^{+51}_{-47}$\\
\smallskip\\\multicolumn{2}{l}{Companion Parameters:}&b\smallskip\\
~~~~$P$ &Period (days) &$117.93116\pm0.00020$\\
~~~~$R_P$ &Radius (\rj) &$1.052^{+0.051}_{-0.049}$\\
~~~~$M_P$ &Mass (\mj) &$79.8^{+3.4}_{-3.3}$\\
~~~~$T_C$ &Time of conjunction (\bjdtdb) &$2455052.70381^{+0.00087}_{-0.00085}$\\
~~~~$a$ &Semi-major axis (AU) &$0.492\pm0.010$\\
~~~~$i$ &Inclination (Degrees) &$89.120^{+0.080}_{-0.076}$\\
~~~~$e$ &Eccentricity  &$0.6859^{+0.0020}_{-0.0021}$\\
~~~~$\omega_*$ &Argument of Periastron (Degrees) &$3.34^{+0.48}_{-0.43}$\\
~~~~$T_{eq}$ &Equilibrium temperature$^{3}$ (K) &$451.5^{+8.5}_{-8.4}$\\
~~~~$\tau_{\rm circ}$ &Tidal circularization timescale (Gyr) &$910000^{+270000}_{-200000}$\\
~~~~$K$ &RV semi-amplitude (m/s) &$4155^{+16}_{-17}$\\
~~~~$R_P/R_*$ &Radius of planet in stellar radii  &$0.09005^{+0.00100}_{-0.0011}$\\
~~~~$a/R_*$ &Semi-major axis in stellar radii  &$88.1^{+3.8}_{-3.4}$\\
~~~~$\delta$ &$\left(R_P/R_*\right)^2$  &$0.00811^{+0.00018}_{-0.00019}$\\
~~~~$\delta_{\rm Kepler}$ &Transit depth in Kepler (fraction) &$0.00860\pm0.00011$\\
~~~~$\tau$ &Ingress/egress transit duration (days) &$0.0373^{+0.0037}_{-0.0035}$\\
~~~~$T_{14}$ &Total transit duration (days) &$0.2520^{+0.0031}_{-0.0030}$\\
~~~~$b$ &Transit Impact parameter  &$0.689^{+0.031}_{-0.036}$\\
~~~~$b_S$ &Eclipse impact parameter  &$0.746^{+0.034}_{-0.040}$\\
~~~~$\tau_S$ &Ingress/egress eclipse duration (days) &$0.0443^{+0.0056}_{-0.0049}$\\
~~~~$T_{S,14}$ &Total eclipse duration (days) &$0.2567^{+0.0022}_{-0.0021}$\\
~~~~$\rho_P$ &Density (cgs) &$84^{+14}_{-11}$\\
~~~~$logg_P$ &Surface gravity  &$5.251^{+0.046}_{-0.043}$\\
~~~~$\Theta$ &Safronov Number  &$69.8^{+3.4}_{-3.2}$\\
~~~~$\fave$ &Incident Flux (\fluxcgs) &$0.00618^{+0.00048}_{-0.00045}$\\
~~~~$T_P$ &Time of Periastron (\bjdtdb) &$2455047.180^{+0.039}_{-0.040}$\\
~~~~$T_S$ &Time of eclipse (\bjdtdb) &$2455040.81^{+0.11}_{-0.12}$\\
~~~~$e\cos{\omega_*}$ &  &$0.6847^{+0.0022}_{-0.0024}$\\
~~~~$e\sin{\omega_*}$ &  &$0.0399^{+0.0056}_{-0.0051}$\\
\smallskip\\\multicolumn{2}{l}{Wavelength Parameters:}&Kepler\smallskip\\
~~~~$u_{1}$ &linear limb-darkening coeff  &$0.356\pm0.042$\\
~~~~$u_{2}$ &quadratic limb-darkening coeff  &$0.285\pm0.048$\\
\smallskip\\\multicolumn{2}{l}{Telescope Parameters:}&APOGEE-N&Keck-HIRES&SOPHIE\smallskip\\
~~~~$\gamma_{\rm rel}$ &Relative RV Offset (m/s) &$6841^{+55}_{-52}$&$-324^{+14}_{-13}$&$6543^{+25}_{-24}$\\
~~~~$\sigma_J$ &RV Jitter (m/s) &$0.00$&$35^{+19}_{-11}$&$16.1^{+4.5}_{-6.8}$\\
\enddata
\label{tab:params_K01288}
\tablenotetext{}{See Table 3 in \citet{Eastman2019} for a detailed description of all parameters}
\tablenotetext{1}{The metallicity of the star at birth}
\tablenotetext{2}{Corresponds to static points in a star's evolutionary history. See \S2 in \citet{Dotter2016}.}
\tablenotetext{3}{Assumes no albedo and perfect redistribution}
\end{deluxetable*}


\newpage
\startlongtable
\begin{deluxetable*}{lcc}
\tabletypesize{\small}
\tablecaption{Median values and 68\% confidence interval for Kepler-849}
\tablehead{\colhead{~~~Parameter} & \colhead{Units} & \multicolumn{1}{c}{Values}}
\startdata
\smallskip\\\multicolumn{2}{l}{Stellar Parameters:}&\smallskip\\
~~~~$M_*$ &Mass (\msun) &$1.259^{+0.12}_{-0.098}$\\
~~~~$R_*$ &Radius (\rsun) &$1.822^{+0.066}_{-0.073}$\\
~~~~$L_*$ &Luminosity (\lsun) &$3.74^{+0.25}_{-0.24}$\\
~~~~$F_{Bol}$ &Bolometric Flux (cgs) &$0.000000000185\pm0.000000000011$\\
~~~~$\rho_*$ &Density (cgs) &$0.292^{+0.055}_{-0.037}$\\
~~~~$\log{g}$ &Surface gravity (cgs) &$4.016^{+0.060}_{-0.049}$\\
~~~~$T_{\rm eff}$ &Effective Temperature (K) &$5950^{+140}_{-130}$\\
~~~~$[{\rm Fe/H}]$ &Metallicity (dex) &$0.145^{+0.059}_{-0.058}$\\
~~~~$[{\rm Fe/H}]_{0}$ &Initial Metallicity$^{1}$  &$0.184^{+0.061}_{-0.059}$\\
~~~~$Age$ &Age (Gyr) &$4.7^{+1.8}_{-1.6}$\\
~~~~$EEP$ &Equal Evolutionary Phase$^{2}$  &$434^{+20}_{-43}$\\
~~~~$A_V$ &V-band extinction (mag) &$0.127^{+0.071}_{-0.080}$\\
~~~~$\sigma_{SED}$ &SED photometry error scaling  &$0.88^{+0.34}_{-0.21}$\\
~~~~$\varpi$ &Parallax (mas) &$1.244^{+0.016}_{-0.017}$\\
~~~~$d$ &Distance (pc) &$803\pm11$\\
~~~~$\dot{\gamma}$ &RV slope$^{3}$ (m/s/day) &$0.0076\pm0.0029$\\
\smallskip\\\multicolumn{2}{l}{Planetary Parameters:}&b\smallskip\\
~~~~$P$ &Period (days) &$394.62508^{+0.00081}_{-0.00079}$\\
~~~~$R_P$ &Radius (\rj) &$0.721^{+0.027}_{-0.029}$\\
~~~~$M_P$ &Mass (\mj) &$0.94\pm0.20$\\
~~~~$T_C$ &Time of conjunction (\bjdtdb) &$2455010.8377\pm0.0016$\\
~~~~$a$ &Semi-major axis (AU) &$1.137^{+0.034}_{-0.030}$\\
~~~~$i$ &Inclination (Degrees) &$89.923^{+0.053}_{-0.058}$\\
~~~~$e$ &Eccentricity  &$0.078^{+0.069}_{-0.053}$\\
~~~~$\omega_*$ &Argument of Periastron (Degrees) &$-74^{+53}_{-61}$\\
~~~~$T_{eq}$ &Equilibrium temperature$^{4}$ (K) &$362.7^{+6.0}_{-5.8}$\\
~~~~$\tau_{\rm circ}$ &Tidal circularization timescale (Gyr) &$2700000000^{+860000000}_{-760000000}$\\
~~~~$K$ &RV semi-amplitude (m/s) &$22.4^{+4.6}_{-4.7}$\\
~~~~$R_P/R_*$ &Radius of planet in stellar radii  &$0.04063^{+0.00040}_{-0.00023}$\\
~~~~$a/R_*$ &Semi-major axis in stellar radii  &$134.0^{+8.0}_{-5.9}$\\
~~~~$\delta$ &$\left(R_P/R_*\right)^2$  &$0.001651^{+0.000032}_{-0.000019}$\\
~~~~$\delta_{\rm Kepler}$ &Transit depth in Kepler (fraction) &$0.002027^{+0.000035}_{-0.000034}$\\
~~~~$\tau$ &Ingress/egress transit duration (days) &$0.0404^{+0.0041}_{-0.0014}$\\
~~~~$T_{14}$ &Total transit duration (days) &$1.0004^{+0.0039}_{-0.0032}$\\
~~~~$b$ &Transit Impact parameter  &$0.19^{+0.15}_{-0.13}$\\
~~~~$b_S$ &Eclipse impact parameter  &$0.17\pm0.12$\\
~~~~$\tau_S$ &Ingress/egress eclipse duration (days) &$0.0368^{+0.0037}_{-0.0043}$\\
~~~~$T_{S,14}$ &Total eclipse duration (days) &$0.905^{+0.094}_{-0.11}$\\
~~~~$\rho_P$ &Density (cgs) &$3.09^{+0.82}_{-0.72}$\\
~~~~$logg_P$ &Surface gravity  &$3.651^{+0.094}_{-0.11}$\\
~~~~$\Theta$ &Safronov Number  &$2.35^{+0.49}_{-0.50}$\\
~~~~$\fave$ &Incident Flux (\fluxcgs) &$0.00389^{+0.00028}_{-0.00027}$\\
~~~~$T_P$ &Time of Periastron (\bjdtdb) &$2454835^{+67}_{-77}$\\
~~~~$T_S$ &Time of eclipse (\bjdtdb) &$2454816^{+15}_{-11}$\\
~~~~$e\cos{\omega_*}$ &  &$0.011^{+0.061}_{-0.045}$\\
~~~~$e\sin{\omega_*}$ &  &$-0.052^{+0.052}_{-0.067}$\\
\smallskip\\\multicolumn{2}{l}{Wavelength Parameters:}&Kepler\smallskip\\
~~~~$u_{1}$ &linear limb-darkening coeff  &$0.385\pm0.030$\\
~~~~$u_{2}$ &quadratic limb-darkening coeff  &$0.276\pm0.045$\\
\smallskip\\\multicolumn{2}{l}{Telescope Parameters:}&Keck-HIRES\smallskip\\
~~~~$\gamma_{\rm rel}$ &Relative RV Offset$^{3}$ (m/s) &$-3.8\pm2.6$\\
~~~~$\sigma_J$ &RV Jitter (m/s) &$8.5^{+2.7}_{-1.9}$\\
\enddata
\label{tab:params_K01439}
\tablenotetext{}{See Table 3 in \citet{Eastman2019} for a detailed description of all parameters}
\tablenotetext{1}{The metallicity of the star at birth}
\tablenotetext{2}{Corresponds to static points in a star's evolutionary history. See \S2 in \citet{Dotter2016}.}
\tablenotetext{3}{Reference epoch = 2458024.297183}
\tablenotetext{4}{Assumes no albedo and perfect redistribution}
\end{deluxetable*}


\newpage
\startlongtable
\begin{deluxetable*}{lccc}
\tabletypesize{\small}
\tablecaption{Median values and 68\% confidence interval for PH-2}
\tablehead{\colhead{~~~Parameter} & \colhead{Units} & \multicolumn{2}{c}{Values}}
\startdata
\smallskip\\\multicolumn{2}{l}{Stellar Parameters:}&\smallskip\\
~~~~$M_*$ &Mass (\msun) &$0.966^{+0.050}_{-0.056}$\\
~~~~$R_*$ &Radius (\rsun) &$0.949^{+0.028}_{-0.027}$\\
~~~~$L_*$ &Luminosity (\lsun) &$0.874^{+0.061}_{-0.050}$\\
~~~~$F_{Bol}$ &Bolometric Flux (cgs) &$0.000000000240^{+0.000000000016}_{-0.000000000014}$\\
~~~~$\rho_*$ &Density (cgs) &$1.60^{+0.16}_{-0.17}$\\
~~~~$\log{g}$ &Surface gravity (cgs) &$4.470^{+0.032}_{-0.038}$\\
~~~~$T_{\rm eff}$ &Effective Temperature (K) &$5734^{+84}_{-82}$\\
~~~~$[{\rm Fe/H}]$ &Metallicity (dex) &$0.017^{+0.051}_{-0.044}$\\
~~~~$[{\rm Fe/H}]_{0}$ &Initial Metallicity$^{1}$  &$0.022^{+0.055}_{-0.053}$\\
~~~~$Age$ &Age (Gyr) &$4.1^{+4.3}_{-2.8}$\\
~~~~$EEP$ &Equal Evolutionary Phase$^{2}$  &$345^{+37}_{-32}$\\
~~~~$A_V$ &V-band extinction (mag) &$0.130^{+0.084}_{-0.077}$\\
~~~~$\sigma_{SED}$ &SED photometry error scaling  &$0.76^{+0.31}_{-0.19}$\\
~~~~$\varpi$ &Parallax (mas) &$2.927\pm0.017$\\
~~~~$d$ &Distance (pc) &$341.6^{+2.0}_{-1.9}$\\
~~~~$\dot{\gamma}$ &RV slope$^{3}$ (m/s/day) &$-0.0104^{+0.0045}_{-0.0050}$\\
\smallskip\\\multicolumn{2}{l}{Planetary Parameters:}&b\smallskip\\
~~~~$P$ &Period (days) &$282.52542\pm0.00011$\\
~~~~$R_P$ &Radius (\rj) &$0.833^{+0.026}_{-0.024}$\\
~~~~$M_P$ &Mass (\mj) &$0.274^{+0.066}_{-0.067}$\\
~~~~$T_C$ &Time of conjunction (\bjdtdb) &$2455196.06770\pm0.00036$\\
~~~~$a$ &Semi-major axis (AU) &$0.833^{+0.014}_{-0.016}$\\
~~~~$i$ &Inclination (Degrees) &$89.898^{+0.021}_{-0.017}$\\
~~~~$e$ &Eccentricity  &$0.215^{+0.095}_{-0.082}$\\
~~~~$\omega_*$ &Argument of Periastron (Degrees) &$24^{+33}_{-16}$\\
~~~~$T_{eq}$ &Equilibrium temperature$^{4}$ (K) &$295.1^{+5.0}_{-4.5}$\\
~~~~$\tau_{\rm circ}$ &Tidal circularization timescale (Gyr) &$46000000^{+29000000}_{-23000000}$\\
~~~~$K$ &RV semi-amplitude (m/s) &$9.0^{+2.1}_{-2.2}$\\
~~~~$R_P/R_*$ &Radius of planet in stellar radii  &$0.09030^{+0.00047}_{-0.00052}$\\
~~~~$a/R_*$ &Semi-major axis in stellar radii  &$188.9^{+6.2}_{-6.8}$\\
~~~~$\delta$ &$\left(R_P/R_*\right)^2$  &$0.008154^{+0.000085}_{-0.000093}$\\
~~~~$\delta_{\rm Kepler}$ &Transit depth in Kepler (fraction) &$0.01001\pm0.00013$\\
~~~~$\tau$ &Ingress/egress transit duration (days) &$0.0406^{+0.0015}_{-0.0016}$\\
~~~~$T_{14}$ &Total transit duration (days) &$0.4506\pm0.0012$\\
~~~~$b$ &Transit Impact parameter  &$0.297^{+0.046}_{-0.063}$\\
~~~~$b_S$ &Eclipse impact parameter  &$0.344^{+0.055}_{-0.069}$\\
~~~~$\tau_S$ &Ingress/egress eclipse duration (days) &$0.0484^{+0.0048}_{-0.0043}$\\
~~~~$T_{S,14}$ &Total eclipse duration (days) &$0.521^{+0.042}_{-0.039}$\\
~~~~$\rho_P$ &Density (cgs) &$0.58^{+0.16}_{-0.15}$\\
~~~~$logg_P$ &Surface gravity  &$2.989^{+0.098}_{-0.13}$\\
~~~~$\Theta$ &Safronov Number  &$0.57\pm0.14$\\
~~~~$\fave$ &Incident Flux (\fluxcgs) &$0.00163^{+0.00012}_{-0.00011}$\\
~~~~$T_P$ &Time of Periastron (\bjdtdb) &$2455162.8^{+14}_{-7.1}$\\
~~~~$T_S$ &Time of eclipse (\bjdtdb) &$2455089^{+19}_{-22}$\\
~~~~$e\cos{\omega_*}$ &  &$0.19^{+0.11}_{-0.12}$\\
~~~~$e\sin{\omega_*}$ &  &$0.080^{+0.042}_{-0.043}$\\
\smallskip\\\multicolumn{2}{l}{Wavelength Parameters:}&Kepler\smallskip\\
~~~~$u_{1}$ &linear limb-darkening coeff  &$0.405\pm0.019$\\
~~~~$u_{2}$ &quadratic limb-darkening coeff  &$0.259\pm0.040$\\
\smallskip\\\multicolumn{2}{l}{Telescope Parameters:}&Keck-HIRES&HARPS-N\smallskip\\
~~~~$\gamma_{\rm rel}$ &Relative RV Offset$^{3}$ (m/s) &$11.8^{+5.5}_{-5.1}$&$-18746.7^{+5.7}_{-6.1}$\\
~~~~$\sigma_J$ &RV Jitter (m/s) &$10.3^{+4.9}_{-3.0}$&$4.7^{+1.7}_{-1.5}$\\
\enddata
\label{tab:params_K03663}
\tablenotetext{}{See Table 3 in \citet{Eastman2019} for a detailed description of all parameters}
\tablenotetext{1}{The metallicity of the star at birth}
\tablenotetext{2}{Corresponds to static points in a star's evolutionary history. See \S2 in \citet{Dotter2016}.}
\tablenotetext{3}{Reference epoch = 2458118.762413}
\tablenotetext{4}{Assumes no albedo and perfect redistribution}
\end{deluxetable*}


\newpage
\section{Bulk Metallicity Modeling Posteriors}

\begin{figure*}[!h]
    \begin{center}
    \begin{tabular}{cc}
        \includegraphics[width=8.5cm]{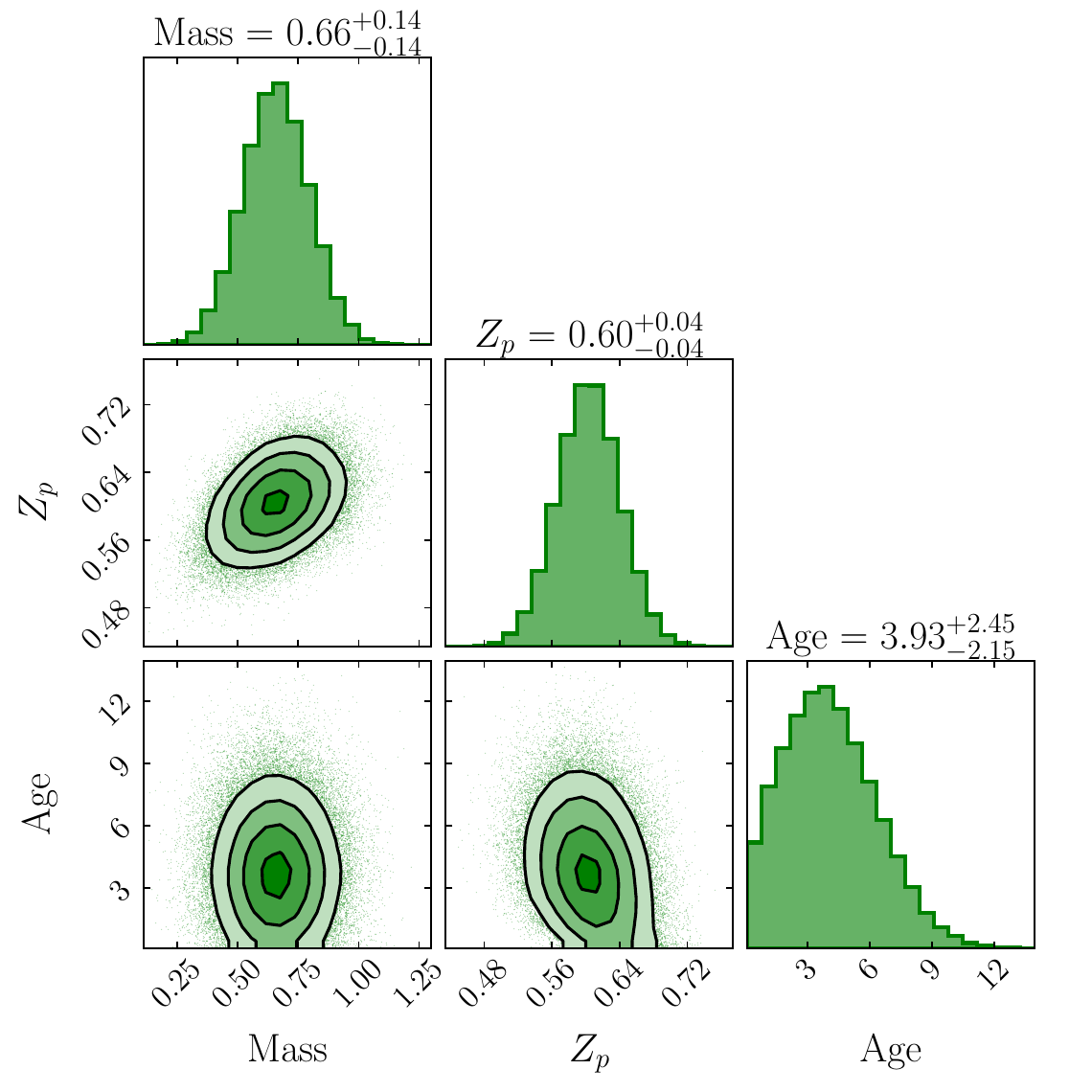} &
        \includegraphics[width=8.5cm]{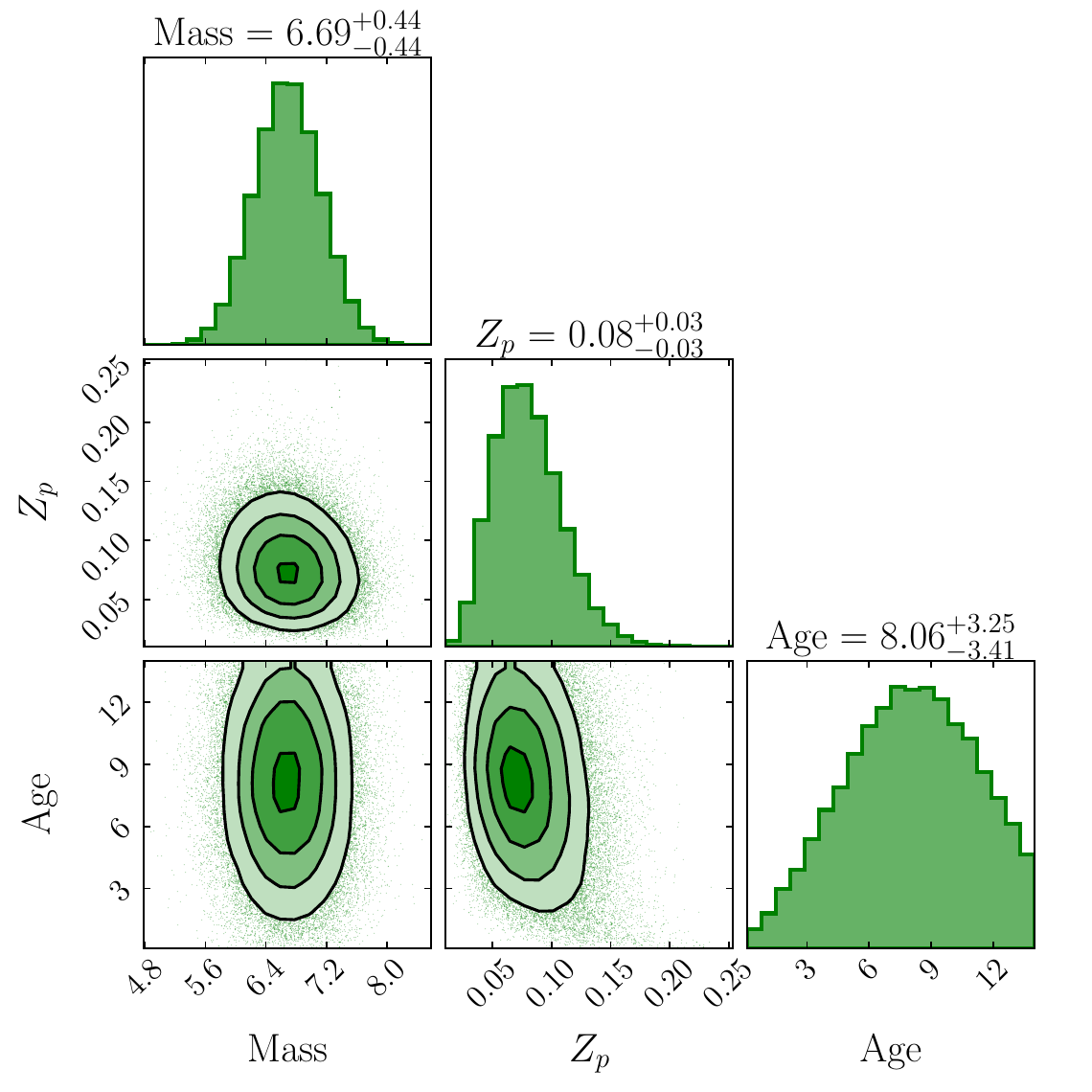} \\
        \includegraphics[width=8.5cm]{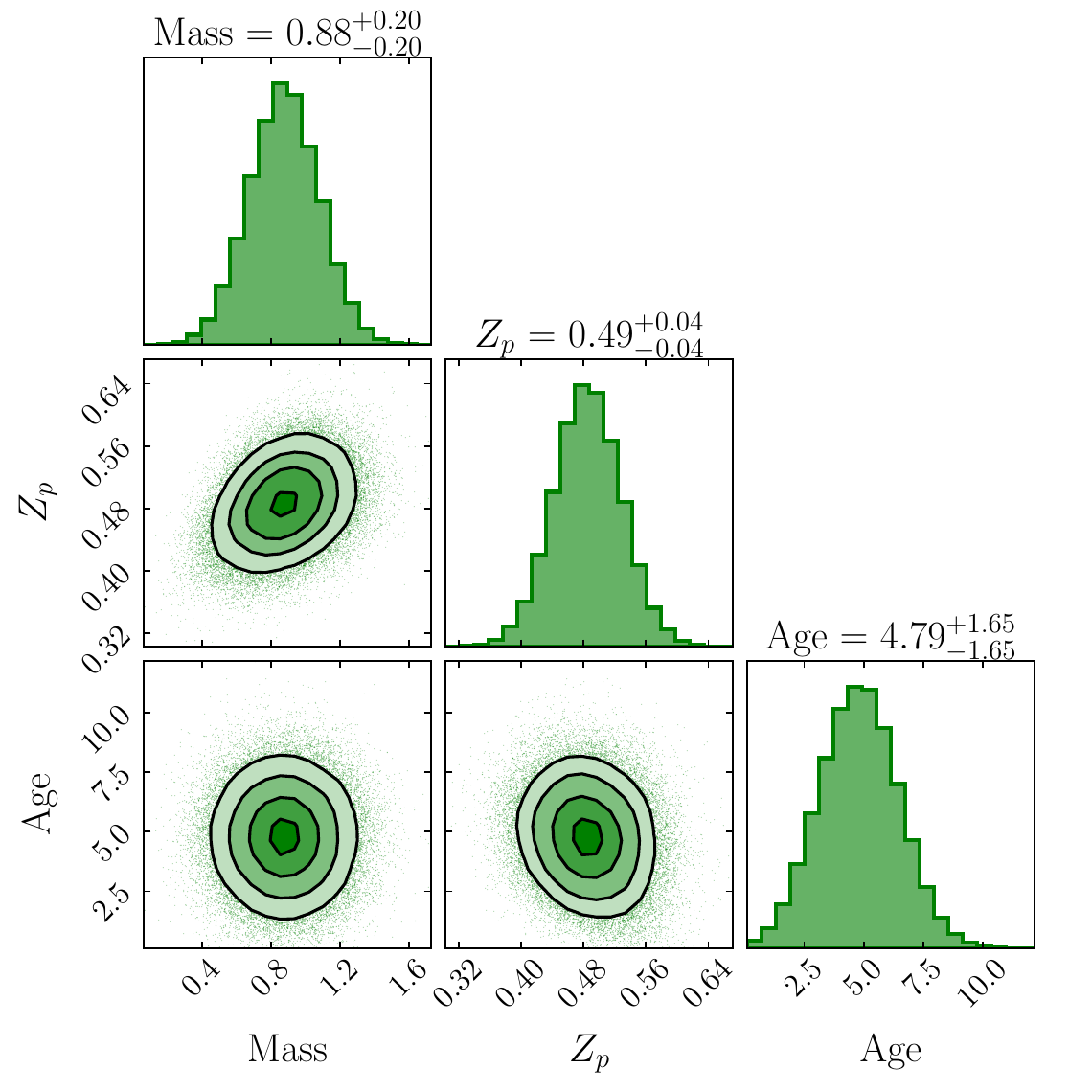} &
        \includegraphics[width=8.5cm]{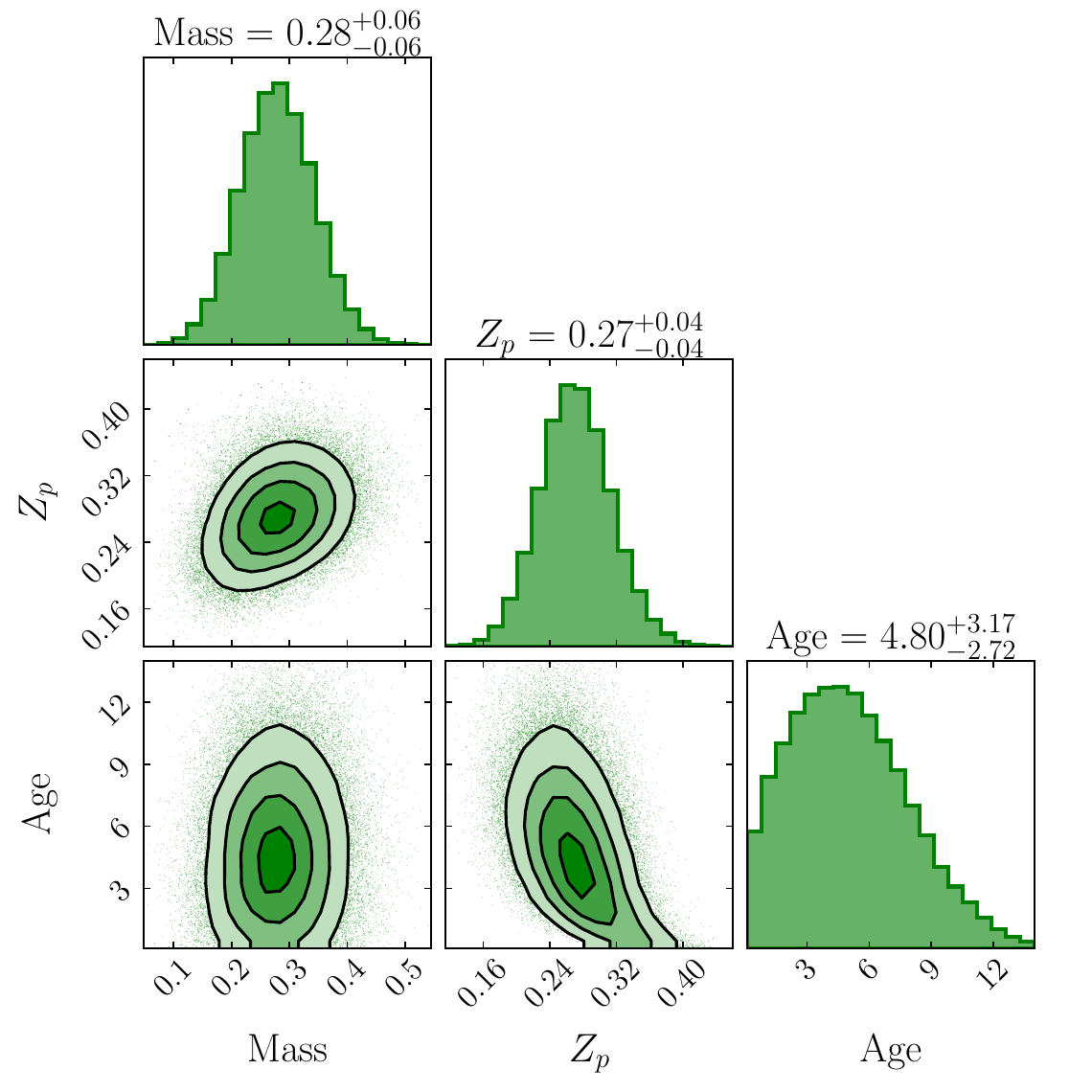}
    \end{tabular}
\end{center}
\caption{Posterior probability distributions resulting from the bulk metallicity estimation described in Section~\ref{sec:metals} for Kepler-111~c (top left), Kepler-553~c (top right), Kepler-849~b (bottom left), and PH-2~b (bottom right). Convergence of the model is achieved in all cases, and high precision is achieved on all of the $Z_P$ values owing to relatively tight constraints on stellar age produced by the \textsf{EXOFASTv2} fits.}
  \label{fig:metals}
\end{figure*}

\newpage
\section{Periodograms of Time Series RV Observations}

\begin{figure*}[!h]
    \centering
    \includegraphics[width=0.85\textwidth]{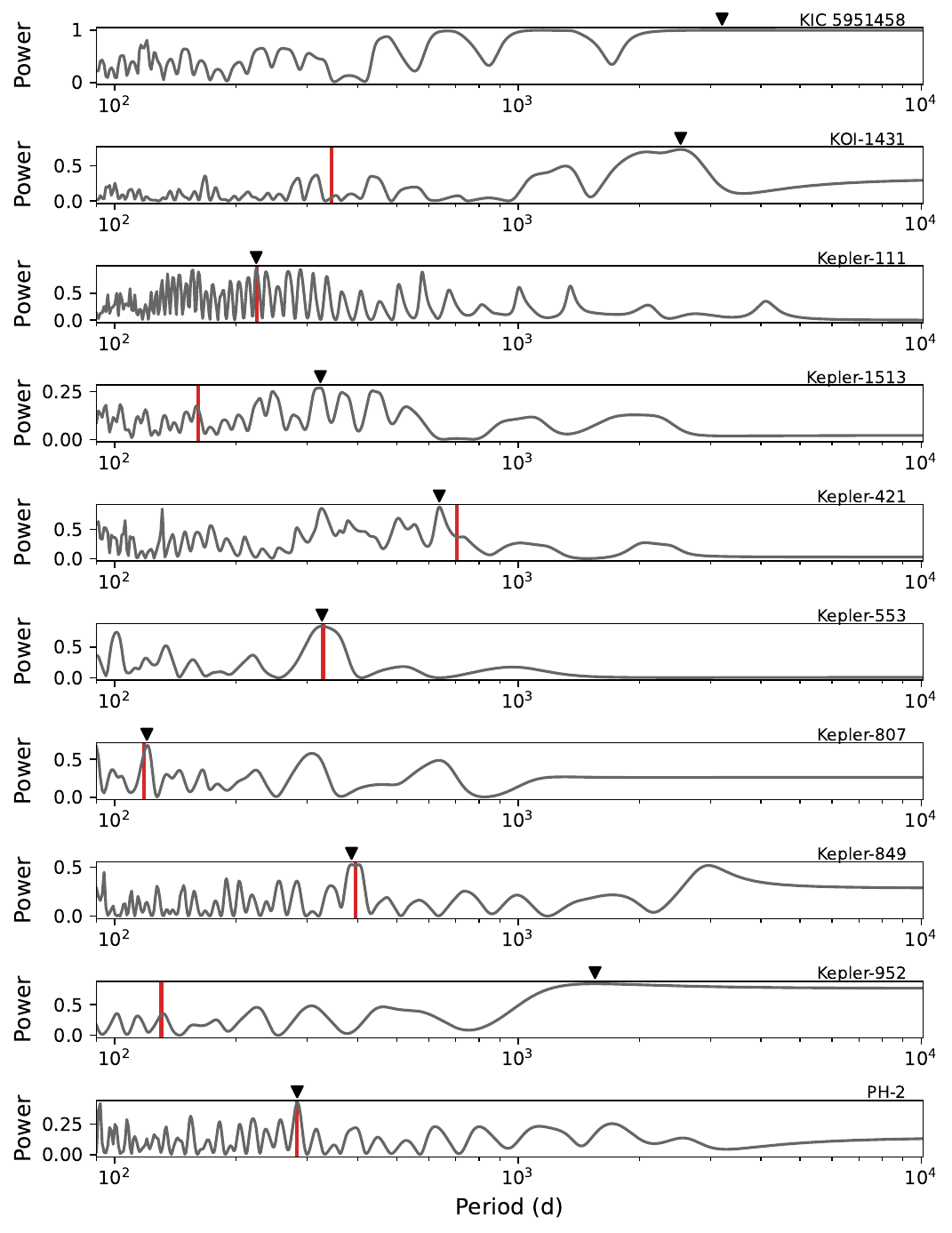}
    \caption{Lomb-Scargle periodograms of the RVs of all systems except for TOI-2180. The red lines identify the periodicity of the transiting giant planet in each of these systems, excluding KIC~5951458, which we argue is spurious. The black triangles show the period with the maximum power in the periodogram. For each of the systems for which we make new mass measurements, these periodicities align.}
    \label{fig:periodogram}
\end{figure*}


\clearpage
\section*{Acknowledgments}
The authors recognize the cultural significance and sanctity that the summit of Maunakea has within the indigenous Hawaiian community. We are deeply grateful to have the opportunity to conduct observations from this mountain. We acknowledge the impact of our presence there and the ongoing efforts to preserve this special place.

We thank the anonymous referee for comments that improved the quality of this research. We thank Jason Eastman for helpful feedback on the usage of EXOFASTv2 and Joseph Rodriguez for helpful discussions related to several of these systems. We thank Ken and Gloria Levy, who supported the construction of the Levy Spectrometer on the Automated Planet Finder. We thank the University of California and Google for supporting Lick Observatory and the UCO staff for their dedicated work scheduling and operating the telescopes of Lick Observatory.

P.D. acknowledges support by a 51 Pegasi b Postdoctoral Fellowship from the Heising-Simons Foundation and by a National Science Foundation (NSF) Astronomy and Astrophysics Postdoctoral Fellowship under award AST-1903811. M.P. gratefully acknowledges NASA award 80NSSC22M0024. J.M.A.M. acknowledges support from the National Science Foundation Graduate Research Fellowship Program under Grant No. DGE-1842400 and from NASA’S Interdisciplinary Consortia for Astrobiology Research (NNH19ZDA001N-ICAR) under award number 19-ICAR19\_2-0041.

Statement of authors' contributions: Authors Dalba through Vowell contributed to the formulation, execution, analysis, and/or management of the GOT `EM survey. Authors Beard through Weiss each took at least 10 of the Keck-HIRES observations published for the first time in this work. Authors Dragomir through Villanueva Jr. assisted in winning the telescope time to support this work. 

Based in part on observations at Kitt Peak National Observatory, NSF's NOIRLab (Prop. ID 2021B-0220; PI: P. Dalba), managed by the Association of Universities for Research in Astronomy (AURA) under a cooperative agreement with the National Science Foundation. The authors are honored to be permitted to conduct astronomical research on Iolkam Du'ag (Kitt Peak), a mountain with particular significance to the Tohono O'odham.

Some of the Keck telescope time used herein was granted by NOAO, through the Telescope System Instrumentation Program (TSIP). TSIP was funded by NSF.

This research has made use of the NASA Exoplanet Archive, which is operated by the California Institute of Technology, under contract with the National Aeronautics and Space Administration under the Exoplanet Exploration Program. This paper includes data collected by the \kepler\ mission and obtained from the MAST data archive at the Space Telescope Science Institute (STScI). Funding for the Kepler mission is provided by the NASA Science Mission Directorate. STScI is operated by the Association of Universities for Research in Astronomy, Inc., under NASA contract NAS 5–26555. This research has made use of the Exoplanet Follow-up Observation Program (ExoFOP; DOI: 10.26134/ExoFOP5) website, which is operated by the California Institute of Technology, under contract with the National Aeronautics and Space Administration under the Exoplanet Exploration Program.

Some of the data presented herein were obtained at the W. M. Keck Observatory, which is operated as a scientific partnership among the California Institute of Technology, the University of California, and NASA. The Observatory was made possible by the generous financial support of the W. M. Keck Foundation. Some of the Keck data were obtained under PI Data awards 2013A and 2013B (M. Payne). 

Finally, P.D. wishes to thank the astronomy community for many years of exciting discoveries and for allowing him the privilege of discovering worlds beyond Earth.


\vspace{5mm}
\facilities{Keck:I (HIRES), APF (Levy), Kepler, TESS}\\

\vspace{5mm}
\software{   \textsf{astropy} \citep{astropy2013,astropy2018},
                \textsf{EXOFASTv2} \citep{Eastman2013,Eastman2017,Eastman2019}, 
                \textsf{lightkurve} \citep{Lightkurve2018}, 
                \textsf{RadVel} \citep{Fulton2018},
                \textsf{ReaMatch} \citep{Kolbl2015},
                \textsf{SpecMatch} \citep{Petigura2015,Petigura2017b}, 
                \textsf{keplerspline} \citep{Vanderburg2016b}}


\end{document}